\documentclass[fleqn,usenatbib]{mnras}

\usepackage{xcolor,subcaption}
\usepackage[export]{adjustbox}
\usepackage{flushend}
\usepackage[utf8]{inputenc}
\usepackage[normalem]{ulem}

\usepackage[compact]{titlesec}  
\titlespacing{\section}{0pt}{12pt}{7pt}
\titlespacing{\subsection}{0pt}{9pt}{4pt}

\newcommand{\PRESTO}{\texttt{PRESTO}}
\newcommand{\flu}{\,Jy\,ms}
\newcommand{\dm}{\,pc\,cm$^{-3}$}

\newcommand\riii{FRB20180916B}
\newcommand\rss{FRB20201124A}
\newcommand\RGC{FRB20200120E}
\newcommand{\rAO}{FRB20121102A}

\usepackage[T1]{fontenc}

\DeclareRobustCommand{\VAN}[3]{#2}
\let\VANthebibliography\thebibliography
\def\thebibliography{\DeclareRobustCommand{\VAN}[3]{##3}\VANthebibliography}

\usepackage{graphicx}	
\usepackage{amsmath}	
\usepackage{amssymb}	
\usepackage{newtxtext,newtxmath}

\title[\rss\ with the uGMRT]{Burst properties of the highly active \rss\  using uGMRT}

\author[Marthi et al.]{V.~R.~Marthi$^1\thanks{Email: \href{mailto:vrmarthi@ncra.tifr.res.in}{vrmarthi@ncra.tifr.res.in}}$,
S.~Bethapudi$^{2}$,
R.~A.~Main$^{2}$,
H.-H.~Lin$^{3,4}$,
L.~G.~Spitler$^{2}$,
R.~S.~Wharton$^{5}$,
D.~Z.~Li$^{6}$, 
\newauthor 
T.~Gautam$^2$,
U.-L.~Pen$^{3,4,7,8,9}$,
G.~H.~Hilmarsson$^{2}$.\\
$^{1}$National Centre for Radio Astrophysics, Tata Institute of Fundamental Research, Post Bag 3, Ganeshkhind, Pune - 411 007, India \\
$^{2}$Max-Planck-Institut f{\"u}r Radioastronomie, Auf dem H{\"u}gel 69, D-53121 Bonn, Germany \\
$^{3}$Institute of Astronomy and Astrophysics, Academia Sinica, Astronomy-Mathematics Building, No. 1, Sec. 4, Roosevelt Road, Taipei 10617, Taiwan\\
$^{4}$Canadian Institute for Theoretical Astrophysics, 60 St. George Street, Toronto, ON M5S 3H8, Canada \\
$^{5}$NASA Postdoctoral Program Fellow, Jet Propulsion Laboratory, California Institute of Technology, Pasadena, CA 91109, USA \\
$^{6}$Cahill Center for Astronomy and Astrophysics, MC 249-17 California Institute of Technology, Pasadena CA 91125, USA \\
$^{7}$Dunlap Institute for Astronomy and Astrophysics, University of Toronto, 50 St. George Street, Toronto, ON M5S 3H4, Canada\\
$^{8}$Program in Cosmology and Gravitation, Canadian Institute for Advanced Research, Toronto, ON M5G 1Z8, Canada\\
$^{9}$Perimeter Institute for Theoretical Physics, 31 Caroline Street North, Waterloo, ON N2L 2Y5, Canada}

\date{Accepted XXX. Received YYY; in original form ZZZ}

\pubyear{2021}

\begin{document}
\label{firstpage}
\pagerange{\pageref{firstpage}--\pageref{lastpage}}
\maketitle

\begin{abstract}
We report the observations of the highly active \rss\ with the upgraded Giant Metrewave Radio Telescope at 550-750~MHz. These observations in the incoherent array mode simultaneously provided an arcsecond localization of bursts from \rss, the discovery of persistent radio emission associated with the host galaxy, and the detection of 48 bursts. Using the brightest burst in the sample ($F= 108~{\rm Jy~ms}$) we find a structure-maximizing dispersion measure of $410.8 \pm 0.5~{\rm pc~cm}^{-3}$. We find that our observations are complete down to a fluence level of $10~{\rm Jy~ms}$, above which the cumulative burst rate scales as a power-law $R(>\!F) = 10~{\rm hr}^{-1} \left(F/10\mathrm{\flu}\right)^{\gamma}$ with $\gamma = -1.2 \pm 0.2$. We find that the bursts are on average wider than those reported for other repeating FRBs. We find that the waiting time between bursts is well approximated by an exponential distribution with a mean of $\sim 2.9$ min during our observations. We searched for periodicities using both a standard Fourier domain method and the Fast Folding Algorithm, but found no significant candidates. We measure bulk spectro-temporal drift rates between $-0.75$ and $-20~{\rm MHz~ms}^{-1}$. Finally, we use the brightest burst to set an upper limit to the scattering time of 11.1~ms at 550~MHz. The localization of \rss\ adds strength to the proof-of-concept method described in our earlier work and serves as a potential model for future localizations and follow-up of repeating FRBs with the uGMRT.


\end{abstract}





\section{INTRODUCTION}

Fast Radio Bursts (FRBs) are short duration (100~$\mu$s-100~ms) 
radio flashes that are extremely bright and appear highly dispersed, and hence thought to be arising from extragalactic distances. Although a 
promising Galactic analogue has been found in SGR~1935+2154 
\citep{chime2020b, bochenek+20}, the origin of these enigmatic 
bursts remains a mystery.  A wide range of FRB models have 
been proposed \citep[see, e.g.,][]{Platts2019}, but none have been conclusively 
proven.  It is not even known yet if there is a single class of 
FRB progenitor or multiple different ones that produce a similar 
observational phenomenon.  


The discovery of repeating FRBs \citep{Spitler2016, scholz2016} marked a 
paradigm shift in FRB research. Repeaters provide an enormous opportunity 
to study individual FRB sources in great detail.  Collecting a large number 
of bursts from a given source can reveal interesting burst structure 
\citep[e.g.,][]{hessels+19}.  Furthermore, since the dispersion measure (DM) 
of the bursts is known for repeaters, data from observations can be coherently dedispersed. This allows us 
to examine the bursts at extremely high time resolution to explore the
emission mechanism and constrain progenitor models \citep{Nimmo+2021, Majid+2021}.  
Repeating FRBs also allow for periodicity searches on a wide range of time scales. 
The discovery of a 16.35-day cycle in the burst activity of \riii\ 
\citep{chime2020a} not only provides an important insight into the progenitor 
source of FRBs, it also allows for highly efficient targeted observations 
\citep[e.g.][]{marthi+20} to be conducted at the precise times at which 
the source is most likely to be active.  A much longer activity cycle has 
also been proposed for \rAO\ \citep{20Rajwade, Cruces+2021}, although 
it remains to be seen if any other repeaters show this cyclic activity. 
Repeating FRBs are also useful targets for searching for short timescale 
periodicities of the order of $\sim$ 1~ms to 1~s that would indicate a 
neutron star origin.  No pulsar-like periodicity has been detected in any 
FRBs \citep{Zhang2018,Li+2021}, but CHIME/FRB recently discovered several 
FRBs with sub-second periodic separations of burst components 
\citep{chime-subsecond-periodicity-2021}.  This result provides further 
evidence in favor of a neutron star origin for FRBs and strongly motivates 
deep searches for pulsar-like periodic emission in repeating FRBs. 


Precise, sub-arcsecond localization of FRBs allows for unambiguous association 
with galaxies and the environs in which they reside. These associations inform 
the choice of the models invoked to explain particular FRBs, as they 
constrain their genesis and evolution, especially through the interplay between 
the progenitor and the circumburst environment 
\citep[e.g.][]{Thompson2017, Margalit2018, Thompson2019}. As an example, the
evolution of the rotation measure (RM) of the repeating \rAO\ has led to 
some interesting constraints on the properties of the circumburst environment
\citep{Hilmarsson2021} and comparison with predictions 
\citep{Piro2018, Margalit2018}. 

On 2021-March-31, the Canadian Hydrogen Intensity Mapping Experiment 
Fast Radio Burst (CHIME/FRB) collaboration reported that \rss\ was in 
a very high burst activity state \citep{CHIMER67ATel}.  Based on this 
report, we proposed for and were allocated Director's Discretionary Time 
(DDT) to observe with the upgraded Giant Metrewave Radio Telescope 
(uGMRT; \citealt{uGMRTpaper}) on 2021-April-5.  Subsequent detections by 
the Commensal Real-time Australian Square Kilometre Array Pathfinder 
(ASKAP) Fast Transients (CRAFT) survey confined the positional uncertainty 
of the source to $\sim 10\arcmin$ \citep{ASKAP-firstATel, ASKAP-secondATel}, 
which meant we could cover the entire uncertainty region with one GMRT 
primary beam at Band-4 (550-750 MHz).  Our observations with the uGMRT, 
along with independent campaigns with ASKAP 
\citep{ASKAP-localization-ATel, ASKAP-lowband-loc-ATel, Fong2021arXiv} 
and the VLA \citep{VLA-localization-ATel, Ravi2021arXiv}, 
localized the FRB to its present coordinates.


This paper is the first in a series of three detailing our studies 
of \rss\ with the uGMRT. Here, we describe the observations and detection 
of 48 bursts from \rss\ and present the burst properties. 
In \citealt{main+2021} (hereafter P-II), we present the first 
ever scintillation timescale measurement of an FRB using combined uGMRT 
and 100-m Effelsberg Radio Telescope observations. 
Finally, in Wharton et al. (2021, \emph{in prep.}; hereafter P-III), 
we detail the precise localization of \rss\ bursts and continuum imaging 
of the host galaxy.  


This paper is organized as follows. Section~\ref{sec:observations}  describes the observations and the telescope configuration, as well as the burst detection and characterization. In Section~\ref{sec:results}, we describe the results of our DM optimization, burst localization, fluence completeness, short timescale periodicity search and the various properties of the bursts, such as the fluence distribution, the distribution of the burst widths, the statistics of the waiting time between bursts and the burst spectral energies. 
    

    


\section{OBSERVATIONS AND DATA}\label{sec:observations}
The uGMRT observations of \rss\ were carried out with the Band-4 receivers 
on 28 of 30 antennas available at the time, tuned to 550-750 MHz, on 2021-April-05 from 
12:30~UTC to 16:30~UTC. The primary beam full width at half maximum (FWHM) 
in this band is  $\sim$ 52\arcmin - 38\arcmin.
At the time, the best localization radius of 10\arcmin\ for \rss\ was provided 
by ASKAP \citep{ASKAP-secondATel}, which is within the 88\% sensitivity contour 
of the GMRT primary beam. We therefore recorded the incoherent array 
(IA; \citealt{uGMRTpaper}) beam.

For the observations reported here, we use the IA beam, in which the voltages from each antenna (and each polarization) are first detected and then added. The root mean square (RMS) noise in the beam is hence $\sqrt{N_A}$ times better than a single antenna. The IA beam has a field of view as large as that of a single GMRT dish. In contrast, the phased array (PA; \citealt{uGMRTpaper}) beam is the sum of the voltages from each antenna (for the two polarizations separately) added in phase before detection. As a result, the RMS noise of the PA beam is $N_A$ times better than that of a single GMRT dish. The size of the beam is hence the same as the synthesized beam of the interferometer. The PA beam is hence $\sqrt{N_A}$ times more sensitive than the IA beam.

The GMRT Wideband Backend (GWB) FX correlator was deployed in the 200-MHz Stokes-I interferometer and 8-bit beamformer mode. An FFT of every 4096 samples of the real-valued voltage time series consumes 10.24~$\mu$s of data. Sixty-four contiguous FFTs are integrated for the two polarizations individually from each antenna to obtain a 2048-channel spectrum every 655.36~$\mu$s, which was adequate for temporally resolving the bursts. The polarizations are finally added in quadrature to obtain Stokes-I beam.

Interferometric visibility spectra were concurrently recorded with the fastest possible integration time of 671~ms to ensure adequate temporal sampling of the dispersed burst pulses, to aid in their imaging. Ideally, the shorter the visibility integration time, the better, as it is crucial for a clear isolation of the bursts and subsequent high fidelity imaging. The radio source 3C138 was used as a flux and phase calibrator due to its relative proximity to the region of pointing, resulting in relatively small slewing overheads. The array phases were tied together on 3C138 before every 40-minute scan on the target. The phase corrections for the antennas, referred to a reference antenna, were estimated from the visibilities. The rephasing scans were also used for flux calibration. A 2-minute test scan on the Crab pulsar served as a control beam for dedispersion. In all we have $\approx$ 180 min of on-target exposure. 



\subsection{Detecting the bursts}\label{ssec:detection}
We performed a standard \PRESTO\footnote{\href{https://github.com/scottransom/presto}{https://github.com/scottransom/presto}}\ \citep*{Ransom2002} search on the incoherent array. The data were visually inspected for radio frequency interference (RFI) and bad channels were manually flagged. The first round of dedispersion was performed with the DM of 414 $\mathrm{pc~cm}^{-3}$ for detecting bright bursts. The data were then dedispersed with a DM of 411\dm, based on visual inspection of the brightest burst. 
The 16-bit unsigned integer (\texttt{uint16}) filterbanks as provided by uGMRT were converted to unsigned 8-bit (\texttt{uint8}) filterbanks, which were searched using standard \PRESTO\ based pipeline. A large amount of RFI caused the down-conversion to be sub-optimal. Therefore, in addition to above, the \texttt{uint16} filterbanks were de-dispersed and searched using \texttt{single\_pulse\_search.py}.
Candidates with a signal-to-noise (S/N) $\geq 8$ (around 6000 in number) were plotted and manually vetted. A large $\rm S/N$ cutoff was used owing to an excess of residual RFI in the data. 
In all we detected 48 bursts over the full exposure time of 180 minutes.


\begin{figure*}
    \centering
    \includegraphics[width=\linewidth,keepaspectratio]{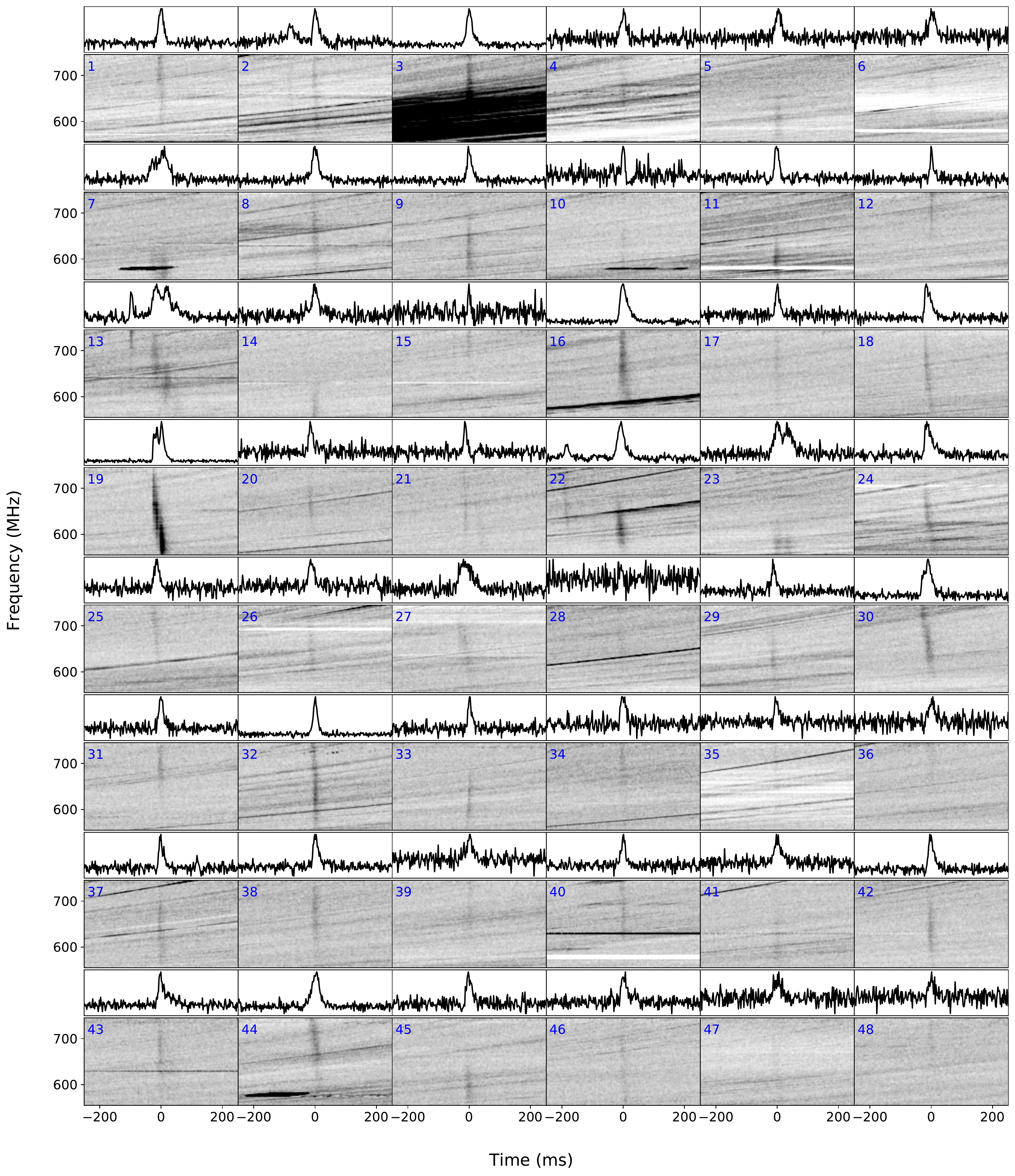}
    
    \caption{Burst profiles and dynamic spectra in chronological order. Each panel shows the dynamic spectrum of a burst from Table~\ref{tab:fluences} dedispersed to 411\dm, with the upper sub-panel showing the frequency-averaged burst profile. The dynamic spectrum has been binned $5\times$ in frequency and $2\times$ in time. The profile, given at the top sub-panel for each burst, is binned $3\times$ in time. We count bursts 2, 13 and 22 as single bursts, owing to difficulty in determining if they are truly different bursts. While it is not apparent in this figure, there is a faint bridge of emission between the components in burst 22, resulting in an end-to-end width of $\sim315$~ms. A similar argument is applied to the components in burst 2, which also is counted as single. The profiles have been cleaned for RFI by subtracting a smooth off-burst baseline through the burst. }
    
    
    \label{fig:burstpanorama}
\end{figure*}

\begin{table*}
    \centering
    \begin{tabular}{r|c|r|r|r|r|r}
    \hline
    Burst  & Barycentric  & Burst width  & Burst width & Peak  & Fluence & Error    \\
       \#  &         TOA  & (zero-crossing) & (boxcar equivalent)  & flux  &     \\
           &   MJD59309+  & ms            & ms &   Jy &   \flu      &  \flu       \\
    \hline
 1 & 0.54335787 & 45.6 & 17.0 &  1.1 &  18.1 &  1.7 \\
 2 & 0.54591884 & 163.8 & 34.3 & 0.8 &  30.1 &  2.4 \\
 3 & 0.54659070 & 91.3 & 23.0 &  1.8 &  41.3 &  1.9 \\
 4 & 0.54761242 & 48.6 & 13.6 & 0.7 &   9.3 &  1.7 \\
 5 & 0.55935229 & 58.5 & 17.8 & 0.7 &  12.8 &  1.7 \\
 6 & 0.57848970 & 65.5 & 14.3 & 0.7 &   9.7 &  1.6 \\
 7 & 0.58136744 & 126.0 & 38.4 & 0.8 &  31.9 &  2.2 \\
 8 & 0.58187302 & 94.3 & 21.4 & 1.2 &  25.8 &  2.5 \\
 9 & 0.58418835 & 57.5 & 16.0 & 1.3 &  20.2 &  1.8 \\
10 & 0.58529781 & 33.7 &  7.5 & 0.6 &   4.8 &  0.8 \\
11 & 0.58921682 & 36.7 &  9.5 & 1.1 &  10.5 &  1.7 \\
12 & 0.59043137 & 33.7 &  10.1 & 0.9 &   9.3 &  1.6 \\
13 & 0.59576209 & 212.3 & 59.3 &  1.0 &  58.4 &  2.9 \\
14 & 0.59626756 & 95.3 & 22.3 & 0.7 &  16.4 &  2.4 \\
15 & 0.59639863 & 19.8 &  6.1 & 0.4 &   2.6 &  0.8 \\
16 & 0.59783325 & 103.2 & 25.2 & 2.1 &  52.9 &  1.8 \\
17 & 0.59870353 & 40.7 & 11.4 & 0.7 &   7.9 &  1.6 \\
18 & 0.61101660 & 69.5 & 19.1 & 1.0 &  19.0 &  1.7 \\
19 & 0.61108629 & 75.4 & 27.9 & 3.9 & 108.0 &  1.6 \\
20 & 0.61256996 & 69.5 & 14.1 & 0.7 &   9.3 &  1.9 \\
21 & 0.61412899 & 33.7 & 7.5 & 0.7 &   5.4 &  1.4 \\
22 & 0.61761505 & 315.5 & 42.1 & 1.6 &  66.5 &  3.8 \\
23 & 0.61763151 & 112.1 & 38.3 & 0.8 &  29.0 &  2.2 \\
24 & 0.62024508 & 79.4 & 23.7 & 0.9 &  21.7 &  2.5 \\
25 & 0.62099280 & 48.6 & 11.9 & 0.8 &   9.0 &  1.5 \\
26 & 0.62315829 & 41.7 & 16.5 & 0.6 &   9.3 &  1.6 \\
27 & 0.62398148 & 86.3 & 25.4 & 0.8 &  20.4 &  2.1 \\
28 & 0.62409520 & 18.9 & 0.7 & 0.0 &   0.0 &  0.1 \\
29 & 0.62932227 & 52.6 & 13.7 & 0.7 &   9.6 &  1.8 \\
30 & 0.63120928 & 84.3 & 22.9 & 1.3 &  29.7 &  1.8 \\
31 & 0.64163927 & 59.5 & 17.2 & 0.7 &  11.9 &  1.6 \\
32 & 0.64281245 & 61.5 & 15.4 & 2.0 &  31.3 &  1.6 \\
33 & 0.64447160 & 51.6 & 13.3 & 0.8 &  10.0 &  1.5 \\
34 & 0.64848111 & 36.7 & 15.3 & 0.6 &   8.9 &  1.6 \\
35 & 0.65177353 & 43.7 & 10.7 & 0.7 &   7.4 &  1.7 \\
36 & 0.65382734 & 44.6 & 15.7 & 0.5 &   7.2 &  1.6 \\
37 & 0.65603984 & 51.6 & 13.1 & 0.9 &  12.1 &  1.7 \\
38 & 0.65638858 & 82.4 & 18.6 & 1.0 &  18.9 &  1.9 \\
39 & 0.65702755 & 68.5 & 15.6 & 0.5 &   8.5 &  1.8 \\
40 & 0.65930498 & 62.5 & 14.8 & 0.8 &  11.2 &  1.6 \\
41 & 0.66312276 & 54.6 & 18.7 & 0.7 &  12.4 &  1.9 \\
42 & 0.66338438 & 63.5 & 16.8 & 1.1 &  17.8 &  1.9 \\
43 & 0.66415983 & 93.3 & 25.1 & 0.9 &  22.2 &  1.8 \\
44 & 0.66577462 & 92.3 & 29.5 & 1.1 &  32.5 &  2.4 \\
45 & 0.67941141 & 45.6 & 15.7 & 0.7 &  11.4 &  1.9 \\
46 & 0.67978340 & 63.5 & 20.9 & 0.7 &  14.7 &  1.9 \\
47 & 0.68965288 & 42.7 & 12.1 & 0.6 &   7.0 &  1.7 \\
48 & 0.68973336 & 53.6 & 10.3 & 0.5 &   5.4 &  1.6 \\
\hline
\end{tabular}
\caption{Time-ordered burst measurements: the ToA is referenced to 550 MHz, the bottom of the band. While burst 28 was detected after RFI cleaning, its fluence could not be estimated reliably.}
\label{tab:fluences}
\end{table*}

\subsection{Flux density calibration}
\label{section:fluxcal}
The detected bursts were isolated for calibration and fluence estimation 
using 1~sec of time-frequency data centered on the dedispersed burst peak. 
First, the mean `off-burst' emission was estimated in every spectral channel, which was then subtracted and divided by, to establish bandpass correction. Similarly, the corresponding phasing scan on 3C138 and an off-calibrator scan were used to measure the deflection on the calibrator. This deflection was scaled appropriately for the spectral shape of the flux density model \citep{Perley2017}. The frequency-averaged time series of the burst and the calibrator scan were then compared to set the flux scale for the bursts, from which we read off the peak flux and estimate the fluence. The time of arrival (ToA) for each burst is measured at the peak, referred to the lowest frequency of the band, 550~MHz. Since the localized position of the FRB  is less than 2.5\arcmin\ away from the antenna boresight (see Section~\ref{ssec:localization}), no primary beam correction has been applied for our fluence estimates, given that the GMRT FoV at Band-4 is $\sim40\arcmin$. The gallery of the detected bursts is shown in Figure~\ref{fig:burstpanorama}. The ToA, burst width, peak flux, fluence and errorbar on the fluence are all listed in Table~\ref{tab:fluences}. 


\section{RESULTS AND PROPERTIES}
\label{sec:results}

\subsection{DM optimization with the power spectrum}\label{ssec:dmopt}
In detecting the bursts, as well as estimating their fluences, widths and ToAs, we use have used DM = 411\dm. However, as we noted in \cite{marthi+20}, the DM that maximizes the peak S/N gives a profile that is degenerate with intrinsic temporal substructure. For bursts that consist of multiple components, a reasonable assumption to consider is that the individual components have no intrinsic drift, while there is only a bulk drift between the components \citep[e.g. in][they make a similar argument]{hessels+19}. One would expect that for the DM that aligns each component perfectly across frequency, it maximizes the total energy in the substructure of a frequency-averaged burst profile. We employ the same method outlined in \cite{marthi+20} to determine the DM that maximizes the energy in the substructure, but we expand on it here for completeness.

\begin{figure}
    \includegraphics[width=\linewidth,keepaspectratio]{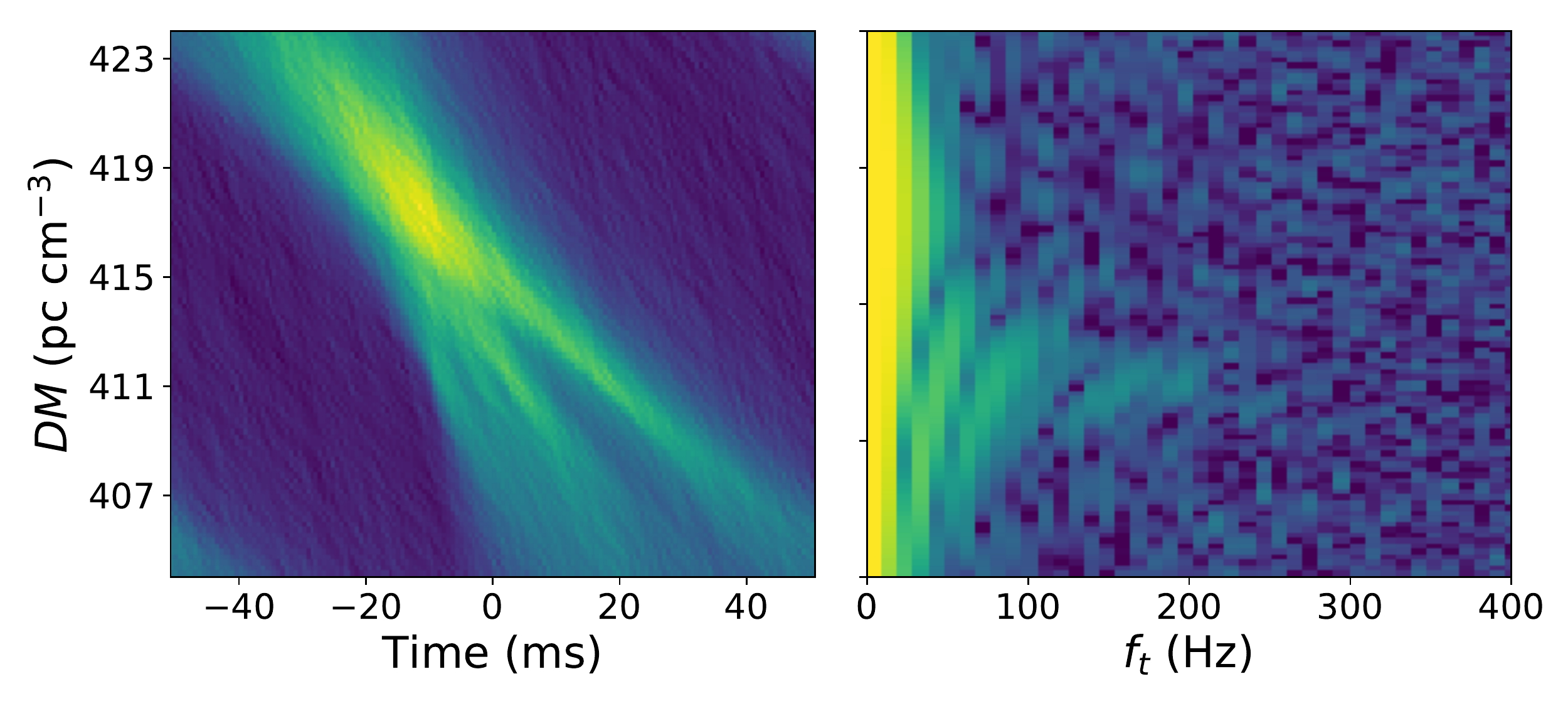}\\
    \includegraphics[width=0.5\linewidth, keepaspectratio]{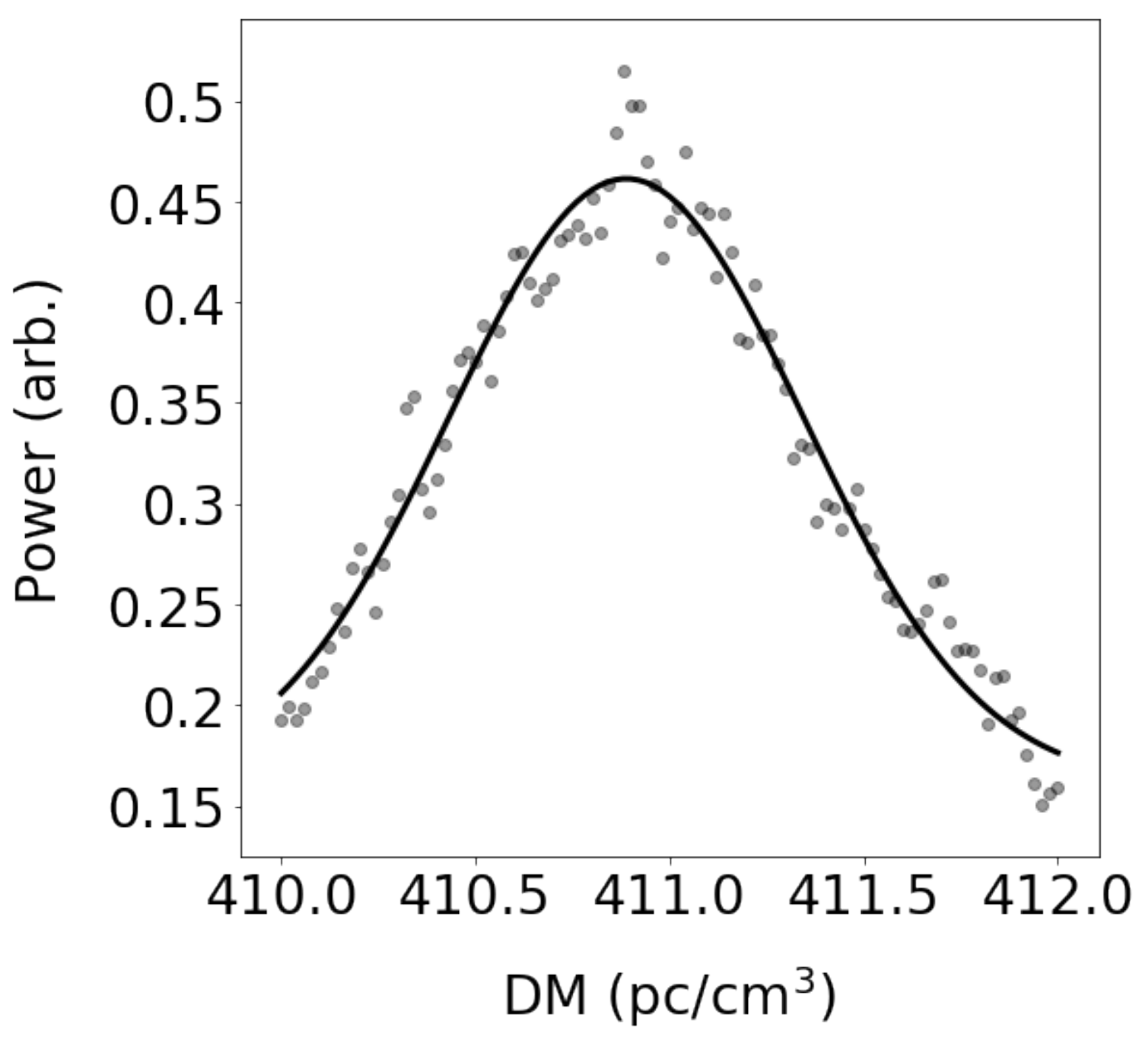}
    \includegraphics[width=0.49\linewidth, keepaspectratio]{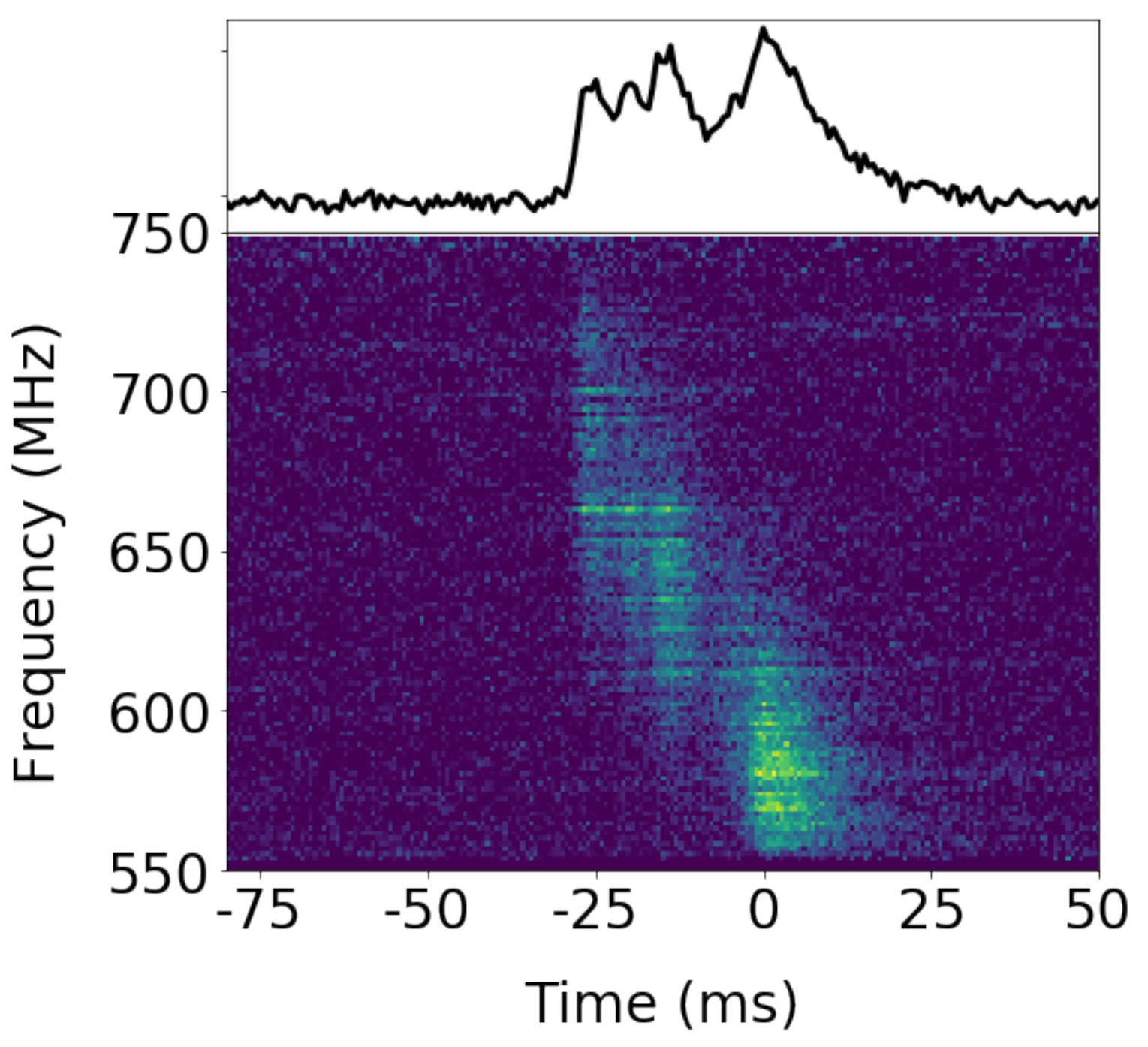}\\
    \caption{DM estimation from burst 19 using the SVD method. \emph{Top Left}:  The frequency-averaged burst profile as a function of DM. Note that the S/N peaks at $\sim$418\dm, but that it compromises the substructure. \emph{Top Right}: The power spectrum as a function of the DM. \emph{Bottom Left}: S/N of the burst power at 180 Hz, following the steps outlined in Section~\ref{ssec:dmopt}. \emph{Bottom Right}: The waterfall and the profile for DM=410.78\dm. } 
    \label{fig:dmfit}
\end{figure}

We incoherently de-disperse the brightest burst for DMs between 410.00\dm\ to 412.00\dm\ in steps of 0.01\dm. The steps undertaken to determine the DM that maximizes substructure energy are as follows: (i) A singular value decomposition (SVD) of the burst dynamic spectrum is taken. The leading mode eigenfunction of frequency for both of the on- and off-pulse, as a weighting function across the frequency, is applied to the dynamic spectrum, and the time-series data determined. (ii) an FFT of the time-series data produces the power spectrum across Doppler frequencies for both of the on- and off-pulse. The power spectrum of the off-pulse is subtracted from that of the on-pulse, giving us the noise-subtracted power spectrum. (iii) the noise-subtracted power spectrum, which we rebinned from 200 Doppler frequencies into 16 in log-scale, is examined from low to high log-Doppler frequencies to determine the cut-off Doppler frequency when the power across DM shows a Gaussian maximum. For the brightest burst, the power versus DM at the cut-off log-Doppler frequency 0.18 kHz is shown in Figure~\ref{fig:dmfit}. (iv) a Guassian profile is fit to the power versus DM to determine the optimal DM and the error, at the peak and the FWHM, respectively. 
The DM thus identified is $410.78\pm0.54 (1\sigma)$ \dm.



We ran {\tt DM\_phase}\footnote{\href{https://github.com/danielemichilli/DM_phase}{https://github.com/danielemichilli/DM\_phase}} \citep{DM_phase_pub}, the DM optimization routine that maximizes the coherent power across the bandwidth, for comparison with our SVD-based method on the brightest burst. It returned a DM of $410.33\pm1.15 (1\sigma)$ \dm, which is in agreement within the error bars with the DM optimized by our method. 


\subsection{Burst localization and persistent radio emission}\label{ssec:localization}
At the time of observation, the best localization \citep{ASKAP-secondATel} for \rss\ had an uncertainty of $15\arcmin$ in right ascension and $5\arcmin$ in declination. This is well within the primary beam of the uGMRT, with excellent prospects for simultaneous localization with the visibilities. Our proposal to observe with the uGMRT focused on the localization of the bursts by using its simultaneous interferometric and beamformer capabilities. We observed with the boresight pointing at (J2000) RA: $05^{\rm h}07^{\rm m}55^{\rm s}$, Dec:$+26\degr02\arcmin00\arcsec$.

After identifying the time of arrival (ToA) of the bursts in the beam, the
visibilities of the brightest burst (Burst 19) were imaged and the burst was 
localized to sub-arcsecond precision \citep{GMRT-localization-ATel}.  
The position was consistent with the earlier VLA localization 
\citep{VLA-localization-ATel} and subsequent localizations with 
the ASKAP low-band \citep{ASKAP-lowband-loc-ATel} and the 
EVN \citep{EVN-localization-ATel}.   
In addition, we detect persistent radio emission of $0.7\pm0.1$~mJy 
at 650~MHz \citep{GMRT-PRS-ATel} coincident with both the burst position and the galaxy 
SDSS J050803.48+260338.0. The persistent radio emission was later found 
to be resolved on milliarcsecond scales \citep{EVN-localization-ATel} 
and likely associated with star formation in the host galaxy 
\citep{Ravi2021arXiv,Fong2021arXiv}.  
The complete analysis and findings of the uGRMT burst localization are 
described in P-III. These observations provide a proof-of-concept for 
future burst localizations of active repeating FRBs with the uGMRT. 


\subsection{Spectral energy}
The most energetic burst in our observations has the largest peak flux. At 108\flu, this bright burst is detected with a S/N of $\sim$40, $\sim 4\times$-$5\times$ lower than ideal due to RFI. This burst would have been detected with a S/N of $\gtrsim$200 with the phased array beam under the same RFI conditions, considering the $\sqrt{N_A}$ boost in sensitivity accrued from co-adding the voltages in phase. 
At a luminosity distance of 451~Mpc (with $h = 0.7, \Omega_\Lambda = 0.7, \Omega_m = 0.3$) \citep[$z=0.098\pm0.02$][]{R67-redshift-Atel, Fong2021arXiv, Ravi2021arXiv}, the isotropic equivalent spectral energy of burst 19 is $2.63\times10^{31}\ \mathrm{erg}\ \mathrm{Hz}^{-1}$. The faintest burst is 2.6\flu, or  $6.33\times10^{29}\ \mathrm{erg}\ \mathrm{Hz}^{-1}$, and is still more than 3 orders of magnitude more energetic than the Galactic FRB SGR~1935+2154 \citep{bochenek+20, chime2020b}. This is roughly two orders of magnitude more than the $10\times$-$25\times$ energy gap for the faintest 0.1\flu\  burst from \riii\ \citep{marthi+20}, but it is not surprising: even for a faint 0.1\flu\ burst that might have been detected with the PA beam, the larger distance of \rss\ means that only those bursts above a threshold spectral energy of $2.43\times10^{28}\ \mathrm{erg}\ \mathrm{Hz}^{-1}$ are automatically selected. We note that the computed spectral energies are lower limits for two reasons: one, often the bursts do not occupy the full band and in addition, there is an unknown beaming factor. The key to bridging the energy gap with the Galactic FRB lies in detecting extremely close FRBs, such as \RGC\ \citep{Bhardwaj2021, Kirsten2021arXiv}.

\subsection{Burst widths}\label{ssec:widths}

We define the boxcar equivalent width of a burst as the ratio of the fluence to the peak flux, measured in ms: this number therefore represents the width of a boxcar window whose height is equal to the peak flux to give the measured fluence. 
The burst widths measured between the zero crossings of a running mean profile, obtained with a 16-bin boxcar kernel, along with the boxcar equivalent width, are given in Table~\ref{tab:fluences}.  However, we use the equivalent width as the basis for simulating events in our completeness analysis in Section~\ref{ssec:completeness}. 

We do not attempt to determine the underlying probability distribution function of the burst widths, as there are no physical models that can inform our choice. Instead, we merely attempt to derive an empirical fit to the data, to aid in the completeness analysis.

\begin{figure}
    \centering
    \includegraphics[width=1.02\linewidth, keepaspectratio]{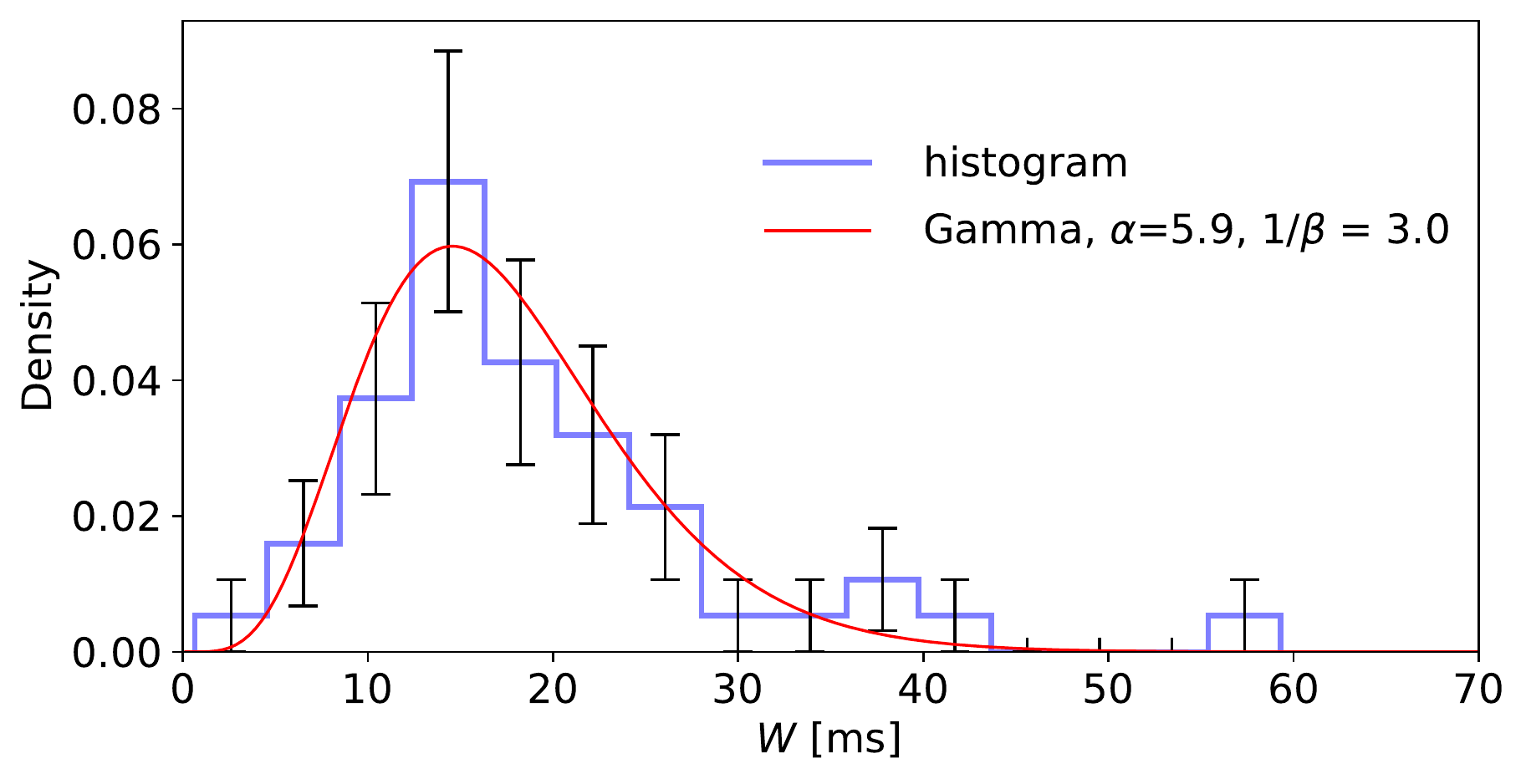}
    \caption{The histogram of the boxcar equivalent burst widths and the PDF of the best-fit Gamma distribution $f_\mathrm{\mathbf{W}}(w; \alpha, \beta)$. The $W>35$~ms bursts are excluded from the fit as they arise from multiple distinct components counted as a single burst. 
        }
    \label{fig:burst-width-PDF}
\end{figure}

Figure~\ref{fig:burst-width-PDF} shows the histogram of the burst width $W$. The errors on the histogram are binwise Poissonian. We find the best fit Gamma distribution for the burst width histogram, which is continuous in the random variable $W$. The Gamma distribution is a general two-parameter family of continuous distributions, of which the exponential, chi-square and Erlang distributions are special cases. The PDF of the Gamma distribution is 
\begin{equation}
    f_W(w; \alpha, \beta) = \frac{\beta^\alpha w^{\alpha -1} e^{-\beta w}}{\Gamma(\alpha)}
    \label{eqn:gamma-PDF}
\end{equation}
where $\alpha$ is the shape parameter and $\beta$ is the rate parameter. The best fit values are $(\alpha, 1/\beta)=(5.9, 3.0)$, obtained after excluding bursts with $W$ > 35~ms. The mean of the PDF is given by $\mu = \alpha/\beta = 17.5$~ms, the standard deviation is $7.2$~ms, and the empirically determined mean width is $\sim$ 17.2~ms. The larger widths correspond to instances where multiple bursts are counted as a single event (see e.g. Figure~\ref{fig:burstpanorama} and Table~\ref{tab:fluences}, burst 2, 13 and 22).


The burst widths of \rss\ are, on average, larger than those from other known repeaters. The majority of bursts from repeaters detected by CHIME/FRB have durations $<$25~ms \citep{2021arXiv210604356P}. We additionally note that the widths of detections of \rss\ at 1.4~GHz are also unusually large \citep{Hilmarsson2021arXiv}. Also, note that since our data are incoherently dedispersed, there is an intra-channel smearing of $\sim2$~ms at the lowest frequency.

\subsection{Fluence distribution}
\begin{figure}
    \centering
    \includegraphics[scale=0.62,trim=0.0cm 0.0cm 0.0cm 0.92cm, clip=True]{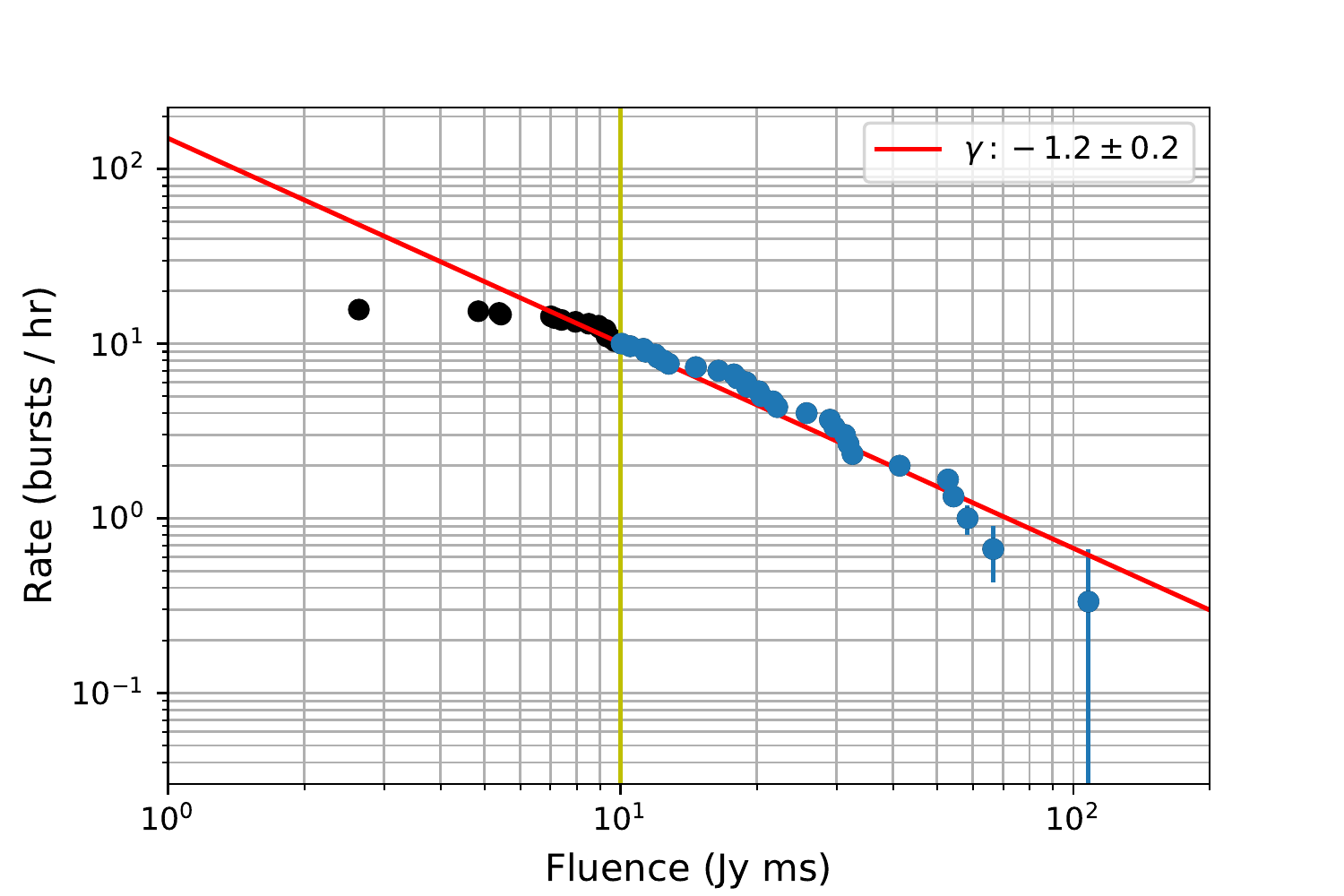}
    \caption{Cumulative burst rate function of the detected bursts. The bursts are complete up to a fluence limit of 10\flu\ (vertical yellow line). The red line is the best-fit power law excluding bursts below the completeness level. The rate function is $R(>\!F) =  10~{\rm hr}^{-1}\left(F/10\mathrm{\flu}\right)^{-1.2}.$ Only the blue data points ($F>$10\flu) are included in the fit, while the black data points are fainter than the completeness limit.}
    \label{fig:fluence-CDF}
\end{figure}



Figure~\ref{fig:fluence-CDF} shows the cumulative burst rate function 
of the burst fluence, $R(>\!F) \propto F^{\gamma}$, which gives the 
rate of bursts above a fluence $F$.  The power-law index, $\gamma$, was 
determined using a maximum likelihood estimator 
\citep[][and references therein]{james+2019} and excluding bursts below 
the completeness limit. Our fit yielded the power-law index of 
$\gamma = -1.2 \pm 0.2$, by setting a visually identified completeness 
limit of 10\flu. We performed a more rigorous analysis with this $\gamma$ 
that gives a completeness limit of 7\flu\ (described in 
Section~\ref{ssec:completeness}). The cumulative burst rate function 
inferred from our observations is

\begin{equation}
R(>\!F) =  10~{\rm hr}^{-1} \left(F/10\mathrm{\flu}\right)^{-1.2}.
\label{eqn:CDF}
\end{equation}


The power law index determined here is shallower than that of the well-studied \rAO\ at 1.4~GHz, where it is a steep $\gamma = -1.8\pm0.3$ \citep{Gourdji2019}, although there is evidence for bi-modality in this source \citep{Li+2021} that can not be well-described by a single power law. CHIME/FRB  determined $\gamma = -1.3 \pm 0.3 \pm 0.1$ for \riii{} at observing frequencies comparable to ours \citep{20R3Period}, which is consistent with our measurement of FRB 20201124A. For comparison, Crab super giant pulses show steeper distributions:  $\gamma =-1.8$ to $-2$ \citep[e.g.][]{ramesh10,Bera2019}. 

Repeating FRBs are subject to strong selection effects on $\gamma$.  Since there are more faint bursts than bright ones, nearby FRBs are more likely to be detected than far away ones.  For $\gamma<-1.5$, the expected distance diverges nearby, meaning we expect to be dominated by nearby FRBs, while for $\gamma>-1.5$ the event rate is dominated by far away ones.  For the observed event rates to converge, we conclude that the intrinsic population has $\gamma>-1.5$ above some luminosity $L_0$, and $\gamma<-1.5$ below.  In a flux limited survey, most FRBs will be detected near $L_0$, and have an apparent $\gamma \sim -1.5$, which is indeed the case for FRB 20201124A.


\subsection{Fluence completeness}\label{ssec:completeness}
For the GMRT incoherent array beam, the RMS noise in 10~ms is $\sigma_\mathrm{IA}\approx30~{\rm mJy}$. For a burst with a peak flux of $S_\mathrm{p}=300~{\rm mJy}$ and pulse width of $W=10~{\rm ms}$, we get a 10$\sigma$ detection:

\begin{equation}
{\rm S/N}_{\rm IA} = 10 \left(\frac{S_{\rm p}}{300~{\rm mJy}}\right)
                              \left(\frac{W}{10~{\rm ms}}\right)^{0.5}  
\end{equation}
For a fluence $F_{\rm p} = S_{\rm p} W$ = 3\flu, 
\begin{equation}
{\rm S/N}_{\rm IA} = 10 \left(\frac{F_{\rm p}}{3~\rm{\flu}}\right)
                              \left(\frac{W}{10~{\rm ms}}\right)^{-0.5}, 
    \label{eqn:detection-likelihood}
\end{equation}
we should be able to detect 10~ms bursts at 10$\sigma$, with brighter, narrower bursts being more likely detected.

The faintest burst we detect in our observations  has a fluence of 2.6\flu, an $\sim 11\sigma$ detection as defined by Equation~\ref{eqn:detection-likelihood}. However, the completeness fluence limit is likely much higher, which we determine as follows.
\begin{figure}
    \centering
    \includegraphics[scale=0.618]{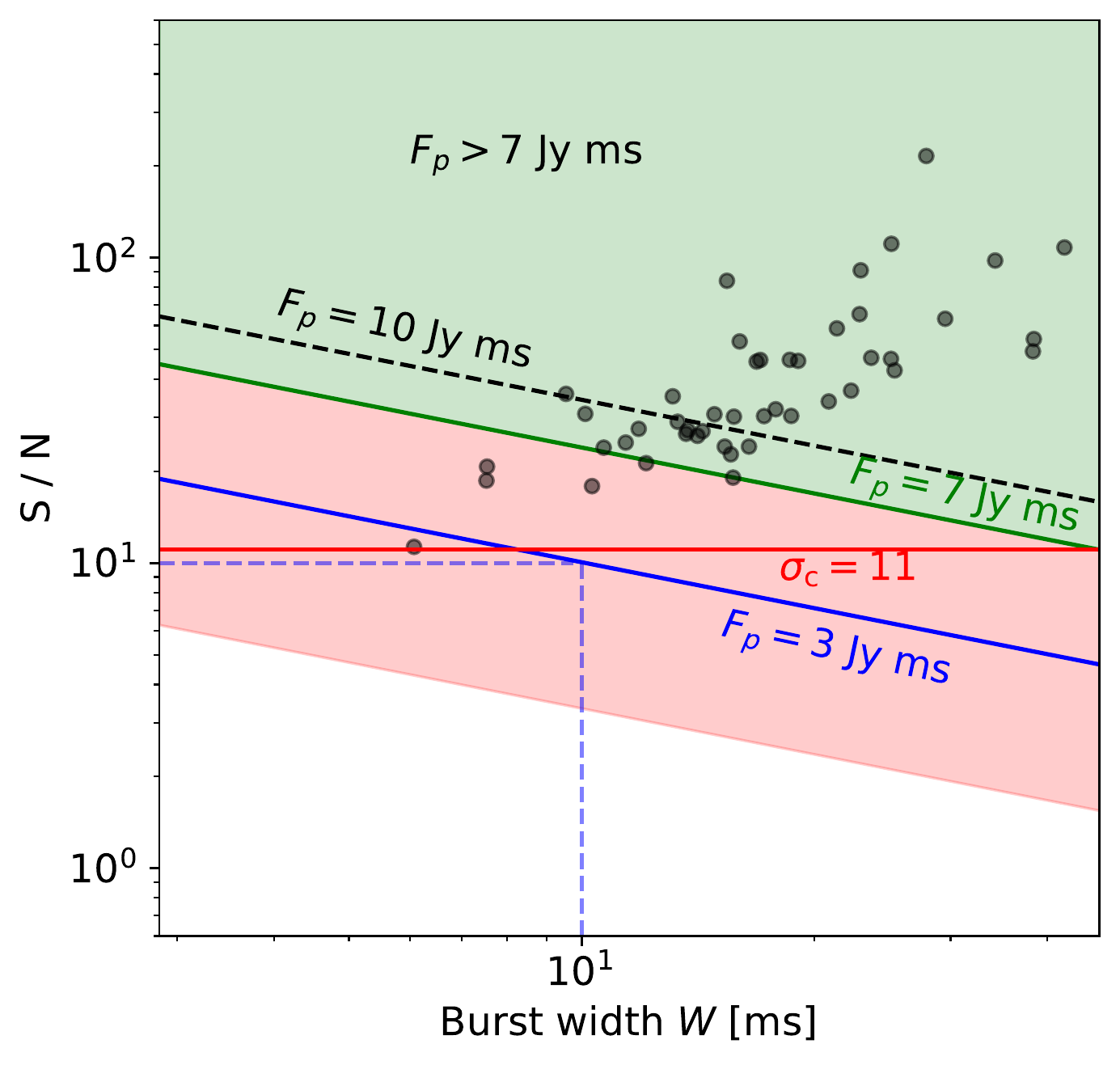}
    \caption{The completeness fluence limit is 7\flu. All bursts that fall in the shaded green region are detected, while only a fraction of those below the limit, in the shaded red region, are detected. The missed fraction is $\sim$80\% in the 1-7\flu\ range denoted by the shaded red region. The $F_p=3$\flu\ line is the reference fluence for a 10$\sigma$ detection of a 10~ms wide burst, according to Equation~\ref{eqn:detection-likelihood}
    The scatter points represent the detected bursts, for which the S/N are computed according to Equation~\ref{eqn:detection-likelihood}. }
    \label{fig:completeness}
\end{figure}

We draw a random sample of $10^3$ bursts from the cumulative fluence distribution of Figure~\ref{fig:fluence-CDF}, using the  empirically determined Equation~\ref{eqn:CDF}.
The lower limit to the fluence is set to 1\flu. Next, we draw a random sample of $10^3$ burst widths given by the Gamma distribution with the best fit ($\alpha, \beta$). Assuming $W$ and $F_p$ are independent, we compute the S/N for every pair (obtained as an outer product) using Equation~\ref{eqn:detection-likelihood}. For a cutoff S/N, defined as $\sigma_c$=(S/N)/$\sigma$=11 (see Figure~\ref{fig:completeness}), we find that the lowest fluence for which bursts of all widths are detected is $\sim7$\flu. 

\begin{figure*}
    \centering
    \begin{minipage}{0.45\linewidth}
    \hskip 0.26in 
    \subcaptionbox{Burst time positions within each scan\label{fig:scaninfo}}{\includegraphics[scale=0.565,trim=0.0cm 0.2cm 0.0cm 0.0cm, clip=True]{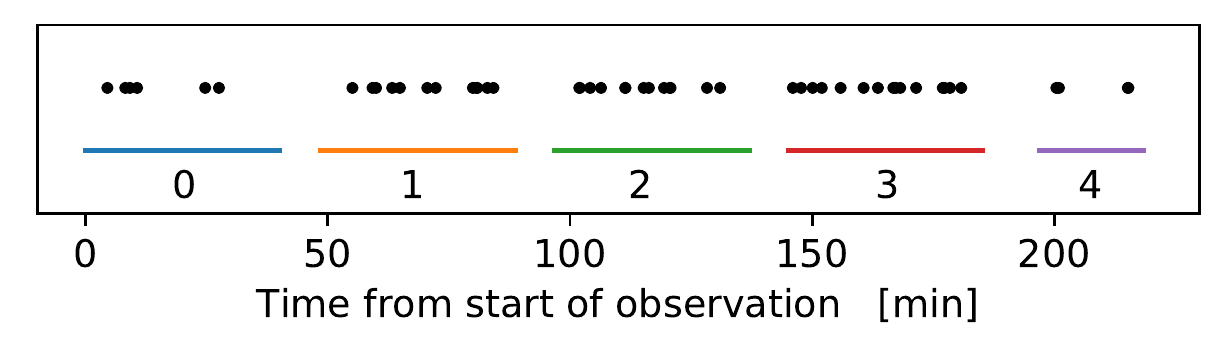}}\\ 
    \subcaptionbox{Burst waiting time\label{fig:wait-time}}{\includegraphics[scale=0.42]{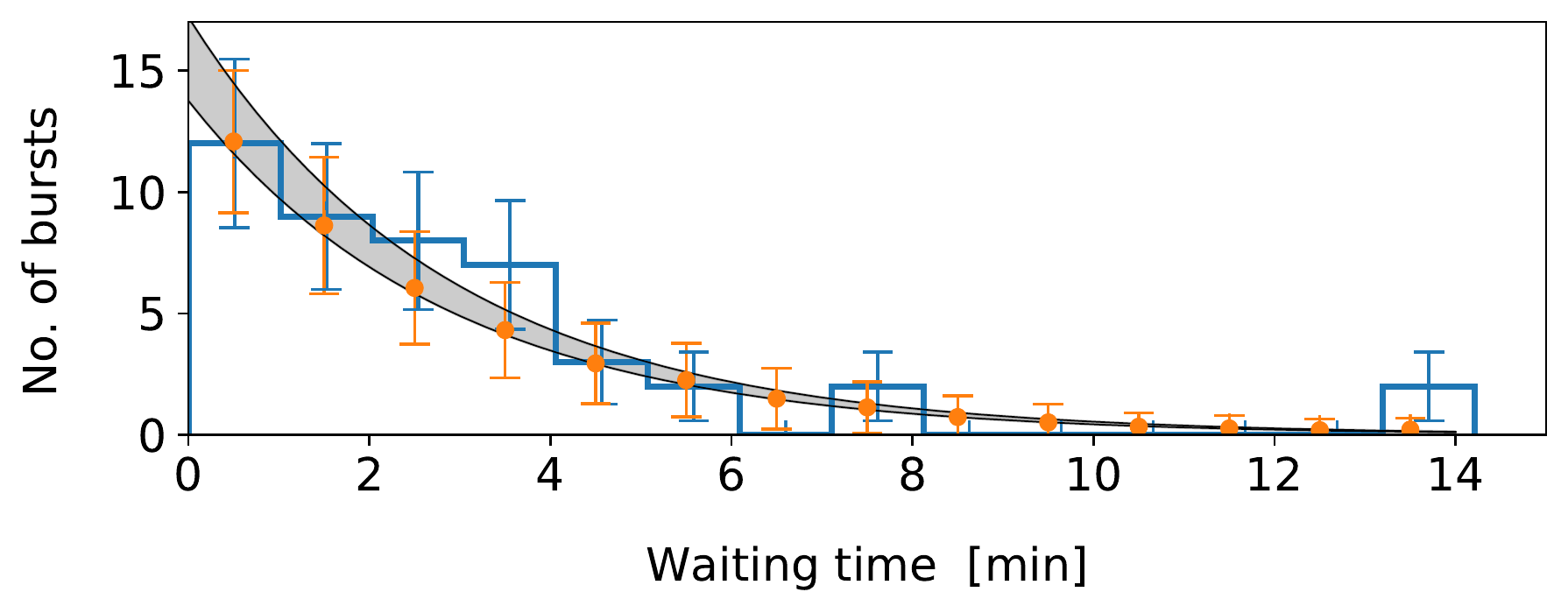} }
    \end{minipage}
    \begin{minipage}{0.45\linewidth}
    \subcaptionbox{Burst pair interval\label{fig:pair-sep}}{\includegraphics[scale=0.4]{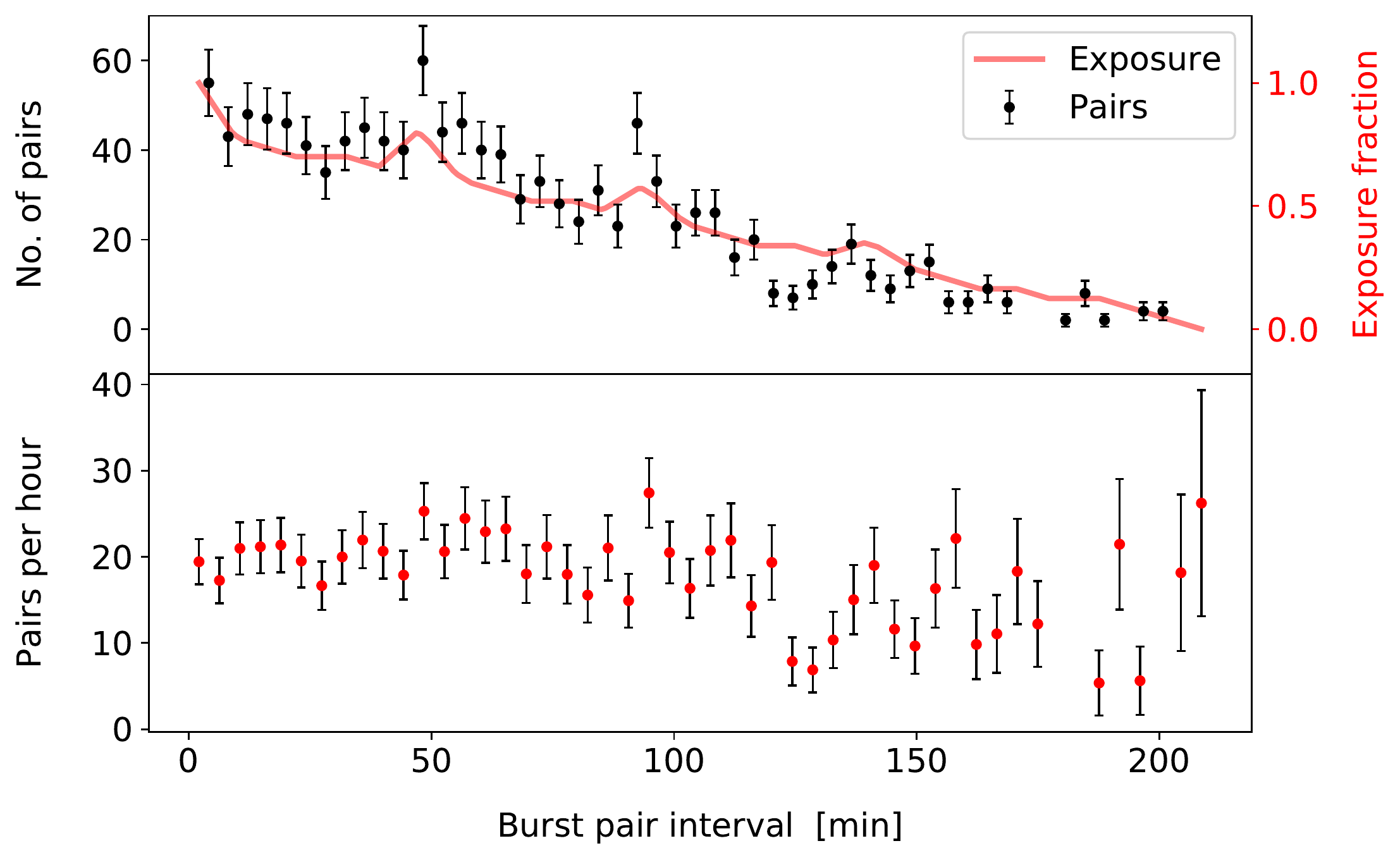} }
    \end{minipage}
    \vspace{2mm}
    \caption{(a): The occurrence of the bursts within each scan. The first four scans are $\approx$40 minutes long each, and the last one is 20 minutes, with 5, 12, 13, 14 and 4 bursts respectively. (b) Histogram of the burst waiting time, with the Poissonian error bars. The histogram of simulated bursts are shown as points with error bars. The shaded region represents the family of exponential mass function curves given by Equation~\ref{eqn:exponPDF} for $40 \leq N_b \leq 50$ with $\lambda = 1/t_w$ (where $t_w = 2.91$ min) from which is drawn the simulated bursts. (c) Burst pair intervals, before (\emph{top}) and after (\emph{bottom}) correcting for relative exposure. The error bars are Poissonian.}
\end{figure*}

We are now specifically interested in the missed fraction of bursts below the completeness limit, up to a reasonable fluence limit. We can compute this quantity as a fraction of the $F_p$ < 7\flu\ events that satisfy the S/N < $\sigma_c$ as well as $F_p > F^\mathrm{th}_p$ condition. For a fluence detection threshold $F^\mathrm{th}_p = 1$\flu, we find that our observations would have missed $\sim80$\% of all bursts 1\flu\ < $F_p$ < 7\flu.
However, our observing conditions were far less than ideal due to the RFI susceptibility of the IA beam. The missed fraction we determine here is hence a lower limit, as we consider 10\flu\ as the more conservative completeness limit.

From the scatter points overlaid on the $W$-S/N space in Figure~\ref{fig:completeness}, it is obvious that bursts with lower widths tend to have lower fluence values. It raises the possibility that the large fraction of bursts being missed could have had intrinsically small burst widths. However, the burst widths follow a Gamma distribution, with the mean burst width being $17.5\pm7.2$~ms, suggesting an inherent deficiency of narrow bursts. On the other hand, it is likely that the burst width distribution itself is biased towards the wider burst population, influenced by the limited sensitivity to narrow bursts (lower fluences).

\subsection{Burst rate and arrival times}
The number of bursts detected in each scan is respectively 5, 12, 13, 14 and 4. The first four exposures are 40 minutes (see Figure~\ref{fig:scaninfo}), while the last one is 20 minutes long.  Considering all bursts above the $\sim11\sigma$ limit of 2.6\flu\ have been detected, we get a burst rate of $\sim$16 per hour, which is likely to be a lower limit due to the incomplete fraction <10\flu. The rate is 10 per hour above the completeness limit of 10\flu. Unlike \riii\ which shows a highly variable burst rate \citep{marthi+20}, \rss\ appears to burst at a more uniform rate at least up to the fluence limit of 2.6\flu. A more sensitive observation with the fully phased beam, which is roughly 5$\times$ more sensitive, could reveal a population of bursts which could revise both the overall burst rate as well as its uniformity. For the PA beam, given an $8\sigma$ threshold of 0.5\flu, the empirically determined CDF of Equation~\ref{eqn:CDF} returns a phenomenal rate of $\sim360$ per hour, if the power law still holds good at the PA beam $8\sigma$ fluence threshold. If confirmed observationally, \rss\ would qualify as the most prolific and active repeating FRB known as yet. \cite{Li+2021} observe a peak rate of 122 hr$^{-1}$ but falling sharply subsequently, suggesting a variable burst rate in \rAO. In \rss, it remains to be seen if the highly optimistic burst rate obtained from an extrapolation of the power law to 0.5\flu\ as well as the uniformity over several epochs hold. It is also likely that our observations occurred at a time when \rss\ was in an extremely active state. Analysis of the cumulative rate distribution from different, widely spaced observations may shed more light on whether the burst rate is evolving.

The longest gap between the scans (see Figure~\ref{fig:scaninfo}) is $\approx$11.7~min, while the typical gap is 8-9~min long. This would skew the histogram of burst wait times slightly, resulting in undercounted bursts for wait times $\lesssim$10~min, while overcounting ones >10~min, assuming a small number of bursts occurred in the scan gaps. In the absence of any priors on the underlying PDF, we cannot account for bursts missed in the rephasing intervals between the scans. One way to circumvent this difficulty is to exclude the pairs that straddle the scan gaps. This results in only a loss of 4 of the 47 waiting times, but results in a more truthful histogram. The mean waiting time between bursts, excluding the 4 inter-scan pairs, is $t_w\sim2.91$ minutes. Figure~\ref{fig:wait-time} shows the histogram of the wait times between successive bursts with the errorbars obtained from the barycentered ToAs, excluding the 4 pairs as described above. 

We ran a simulation by drawing waiting times from an exponential mass function with the empirically determined mean waiting time $t_w$, 
\begin{equation}
    f(t;t_w) = N_b \frac{1}{t_w} e^{-(t/t_w)}, 
    \label{eqn:exponPDF}
\end{equation}
generating $N_b = 50$ bursts in each run. These bursts were distributed within the observation as if they were observed with the same scan durations and intervals shown in Figure~\ref{fig:scaninfo}. Those bursts which fell in the scan intervals, as well as at the book-end bursts in each scan, were excluded before obtaining the histogram. The mean and the errorbar in each bin was determined from 1000 iterations: however, increasing the number of iterations beyond $\sim50$ has very little effect on the mean and the Poissonian errorbars as they tend to converge. We note that the simulated waiting times drawn from an exponential distribution agree well with the burst waiting time histogram, suggesting a good match of the data with the distribution. The outputs from the simulation are plotted as points with errorbars. Curves for the exponential mass function are overplotted for $40 \leq N_b \leq 50$ (see Figure~\ref{fig:wait-time}).

An exponential distribution for the waiting time suggests an underlying Poisson point process with a rate parameter $\lambda = 1/t_w$. This is expected for bursts detected within a single observation, as seen for \rAO\ \citep{Cruces+2021}, which is known to follow a Weibull distribution over a longer time with a clustering factor of $k=0.34^{+0.06}_{-0.05}$ \citep{Oppermann2018}. With only 48 bursts, of which only a fraction is complete (30 of 48 bursts, above 10\flu), there may be pitfalls to modelling the burst waiting time distribution. At this juncture, we limit our discussion to the empirical fit of the distribution. The statistics should benefit significantly from sensitive PA beam detections with a much deeper completeness limit. In addition, continued monitoring over long periods should assist in identifying temporal clustering behaviour if any, considering that it has recently entered a state of heightened activity.



\cite{Li+2021} report waiting time statistics for \rAO\ using a collection of $\sim$1600 bursts. They find the waiting time between bursts to be well fit by a log-normal distribution. Their waiting time histogram shows two peaks, with the largest mean waiting time being $\sim 70\pm12$~s, which they consider to be an upper limit to any periodicity, and a separate $\sim 220\pm100$~s limit to the periodicity of the high energy ($ > 3 \times 10^{38}$ erg) bursts. They conclude that these values are consistent with the mean values for the respective samples, and therefore consistent with underlying stochasticity \citep{Li+2021}. While we do find a few bursts with multiple components separated by $\sim$90~s or $\sim$180~s, we do not have sufficient sensitivity to identify them as distinct and independent bursts. Such closely separated bursts deserve keen attention to determine if they truly arise from a different distribution peaking at a shorter mean waiting time, such as seen for \rAO\ by \cite{Li+2021}.

In addition, we derive a histogram of the pairwise intervals between all burst pairs, to investigate any trends for a preferred separation time, or multiples thereof, between bursts that might hint at longer periodicity timescales or burst rate modulation. The raw histogram is
shown in the top panel of Figure~\ref{fig:pair-sep}. This, however, is biased by the relative exposure between pairs. The relative exposure as a function of burst pair separation is obtained as the autocorrelation function (ACF) of the scan windows (see Figure~\ref{fig:scaninfo}), scaled for the number of burst pairs at zero separation (which is the total number of detected bursts). Obviously, shorter pair separations have a higher relative exposure than longer ones. The peaks in the relative exposure correspond to the $\sim45$-minute interval between successive scan start times. The normalized histogram is given in the bottom panel. We find no evidence for a preferred pair separation time that might indicate longer periodicities or rate modulation of $\sim$minutes timescales.


\subsection{Spectro-temporal drift}
The incoherent array beam is very susceptible to radio frequency 
interference (RFI).  Broadband, zero DM RFI bursts show a characteristic 
reverse dispersion sweep after dedispersion. A singular value decomposition 
of the contaminated, but dedispersed, burst dynamic spectrum breaks the 
reverse-swept RFI features and distributes it across a very large number 
of modes. While the total energy in the RFI is very high, often much more 
than the burst itself, it does not dominate the rank-ordered singular values. 
The dominant orthogonal modes of the dedispersed burst dynamic spectrum can  
thus purely represent the burst. This filtering method works well for bursts 
with reasonably high S/N, but becomes progressively less effective with 
deteriorating S/N. This is not surprising, as the ability to exclude the 
reverse-swept RFI modes depends on their rank ordering in the presence of 
the compact representation of the true burst modes.
   
\begin{figure}
    \includegraphics[scale=0.445]{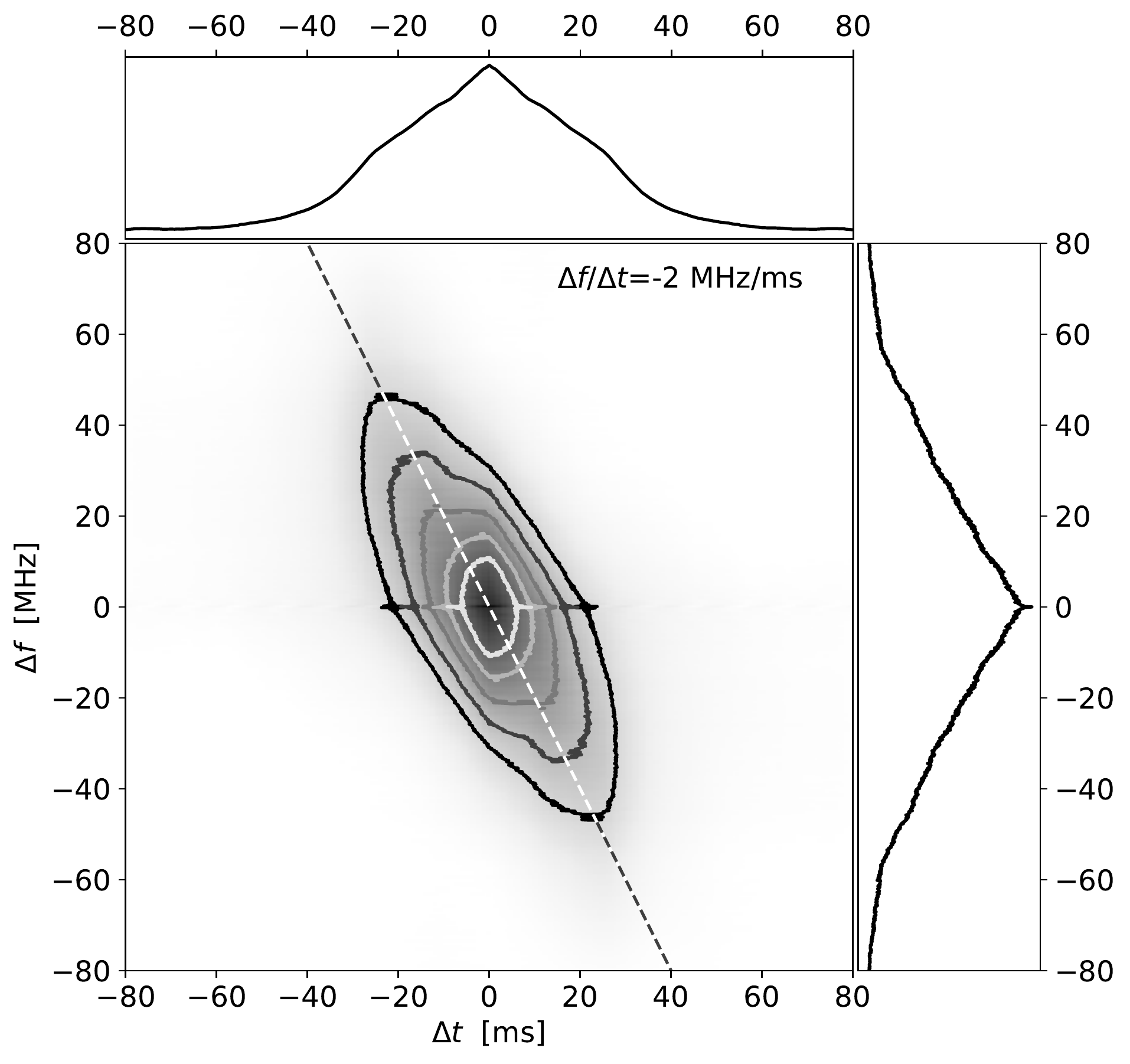}
    \vspace{-3mm}
    \caption{2D ACF of burst 19, used for measuring the drift rate. The upper and the right panels show respectively the frequency- and time-averaged 2D ACF. }
    \label{fig:ACF}
\end{figure}

As an example, in Figure~\ref{fig:ACF} we show the 2D ACF and the estimated drift rate for the brightest burst, as it has more than twice the S/N of any other burst. For this burst (burst 19), we find that, using the SVD mode filtering described above, the first six modes adequately represent the burst, allowing us to exclude all the subsequent modes and thus nearly fully eliminate the swept broadband RFI features. The reconstructed dynamic spectrum was used to obtain the 2D ACF, allowing us to estimate a drift rate of $\Delta f/\Delta t = -2\ \mathrm{MHz}\  \mathrm{ms}^{-1}$. Other bursts show drift rates between $\sim-0.75~{\rm MHz~ms}^{-1}$ and $\sim-20~{\rm MHz~ms}^{-1}$.

All the bursts in our sample show a downward drift or a ``sad trombone'' effect, as is known for repeaters \citep{hessels+19, Caleb2020, Chawla2020}, although some anomalous drifting behaviour has been seen in \riii\ \citep{Pleunis2021}. In \cite{marthi+20}, we see two instances of potential upward (or positive) drift in \riii, but it is not clear if these are from the same or separate bursts. \cite{Platts2021} consider the possibility of such upward drift arising from lensing events in \rAO.

We defer an analysis of the drift-rate vs burst width relation to a future paper, after including more sensitive burst detections. \cite{Chamma2020arXiv}   posit the linear relation between the quantities as a possible universal relationship for repeating FRBs. \cite{Hilmarsson2021arXiv} find the relationship between drift rates and burst widths in 1.4~GHz Effelsberg Radio Telescope observations to be consistent with the results of \cite{Chamma2020arXiv} except for an offset. The primary constraint for a similar but robust analysis of the sample presented in this paper is the unfavourable RFI contamination of the bursts, especially the weaker ones.

   
\subsection{Scattering and scintillation}

\begin{figure*}
    \centering
     \includegraphics[width=0.325\linewidth, keepaspectratio]{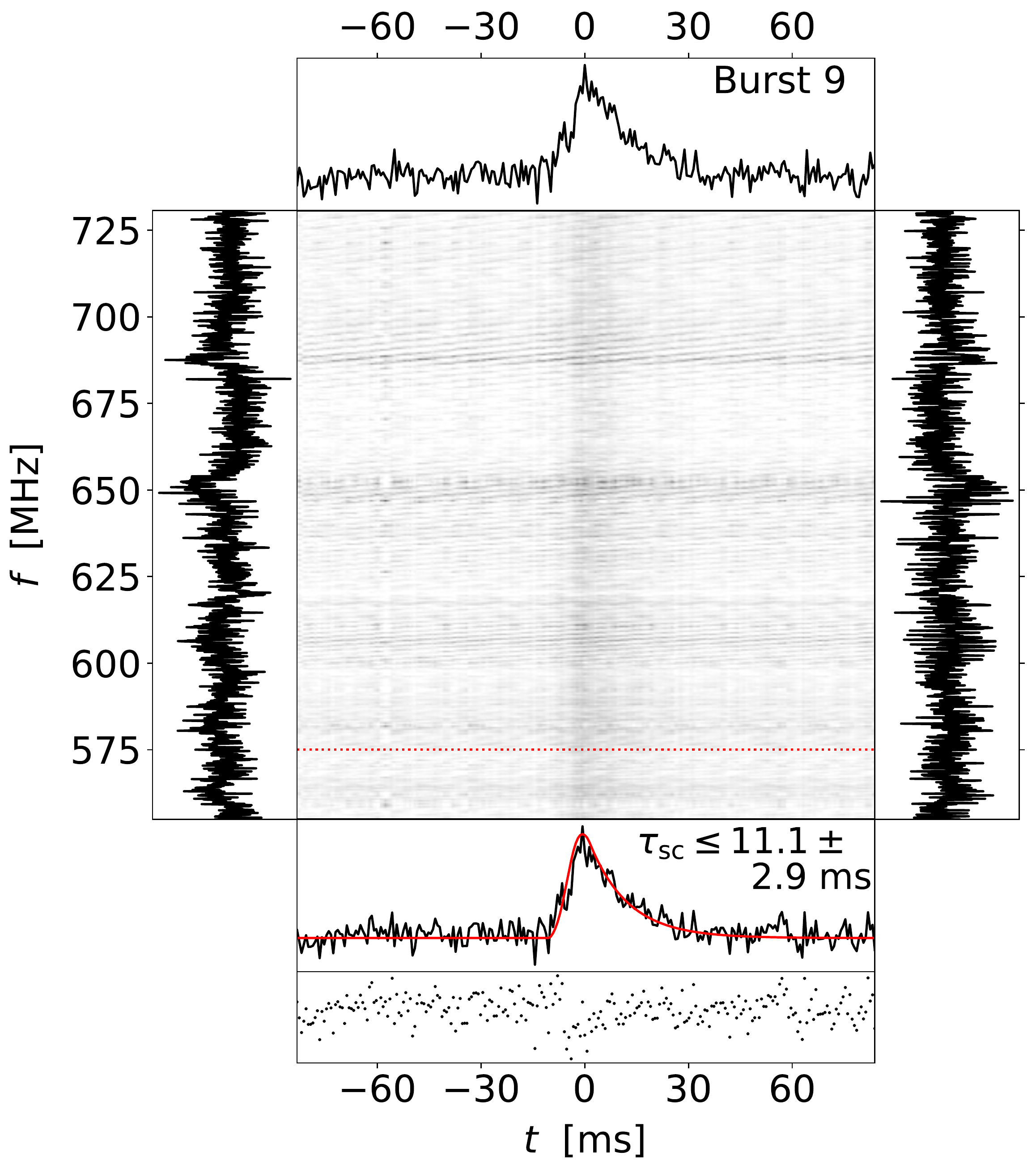}
     \includegraphics[width=0.331\linewidth, keepaspectratio]{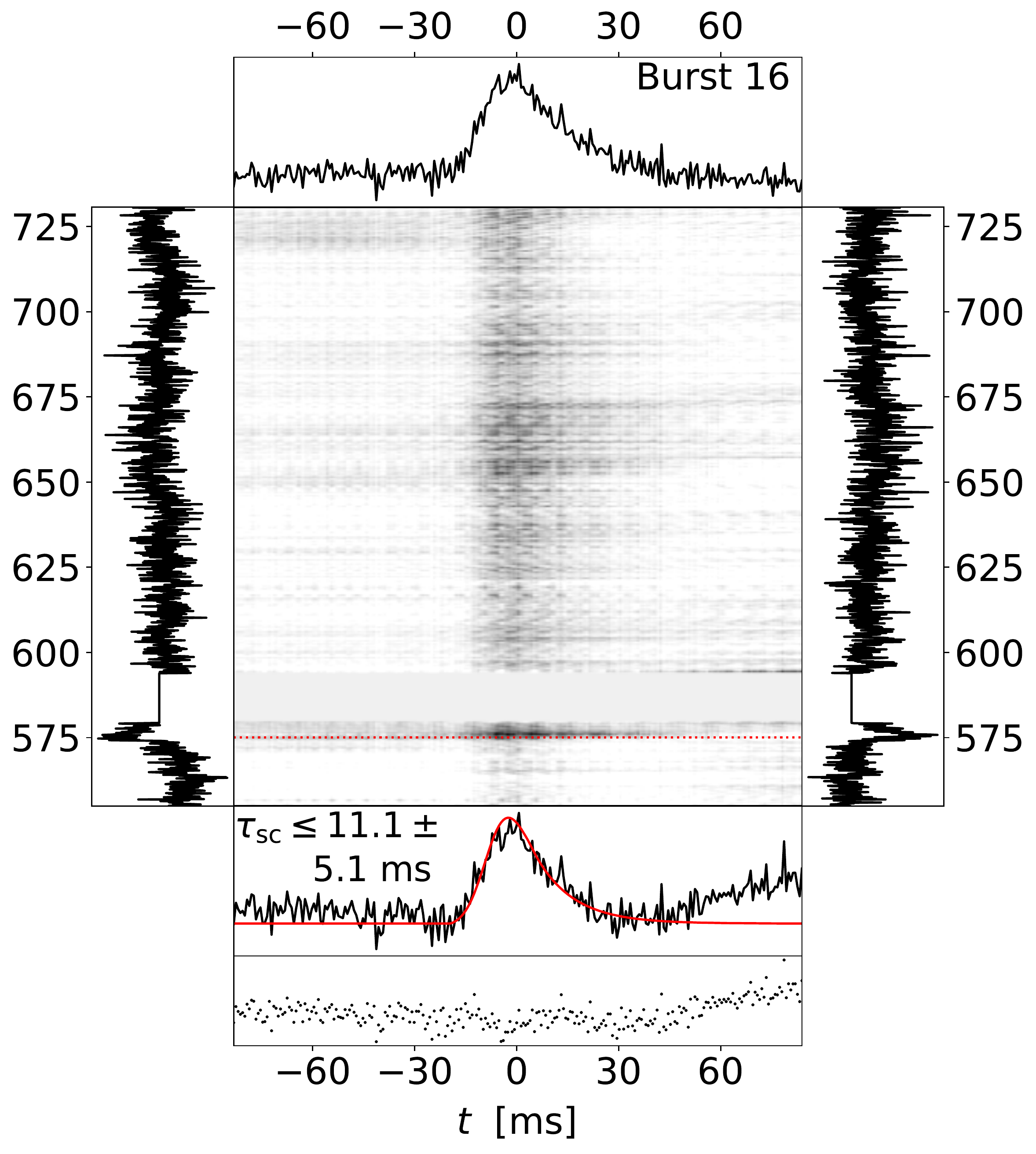}
     \includegraphics[width=0.325\linewidth, keepaspectratio]{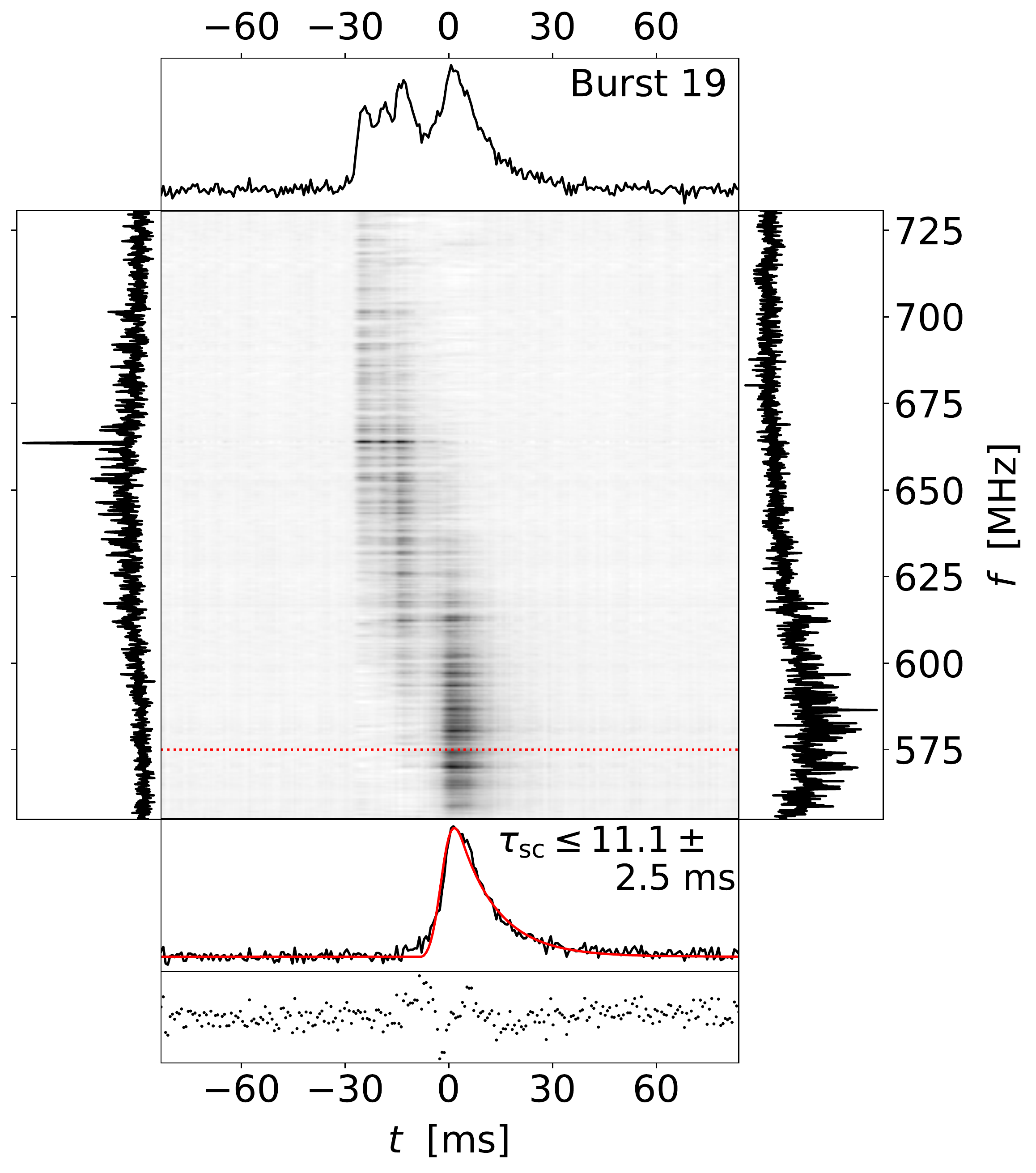}
    \caption{The SVD-filtered dynamic spectra of bursts 9, 16 and 19, dedispersed to 411\dm\ and binned 4$\times$ in frequency for plotting. For all the three bursts, the top panel shows the mean profile integrated between 554~MHz and 731~MHz. The bottom panel shows the low-band profile cut off at 575~MHz. The red curve is the best fit obtained by convolving a Gaussian with a decaying exponential, giving a scattering time constant of 11.1~ms along with the respective 1$\sigma$ error bar, which are 2.9~ms, 5.1~ms and 2.5~ms respectively for bursts 9, 16 and 19. Below this panel are the residuals of the fit. The left and right sub-panels respectively show the burst spectrum at $t<0$ and $t>0$.}
    \label{fig:diagnostics}
\end{figure*}

Figure~\ref{fig:diagnostics} shows the dynamic spectrum of burst 9 on the left, burst 16 in the middle and that of the brightest burst (19) on the right. These bursts have a clearly identifiable decaying exponential profile: we choose bursts 9 and 16 particularly for their apparent lack of spectro-temporal drift. We consider the low frequency portion of the bursts, where a single component of the multi-component burst 19 is isolated by defining a cut-off frequency (here, 575~MHz). The other two bursts have a single component, but the cut-off frequency is defined and the analysis is done identically for consistency. The bottom panel shows the frequency-averaged (550-575~MHz) profile, which shows 
what appears to be an exponential tail. We fit to all the three bursts a Gaussian convolved with a decaying exponential, but constraining the decay time constant to be identical. If we interpret this as a scattering tail, we can set an upper limit of $\tau_\mathrm{sc} \leq 11.1$~ms  to the scattering time. The scattering time $\tau_\mathrm{s}$ is measured as the time constant of the decaying exponential $e^{-t/\tau_s}$, which includes the uncorrected 2-ms dispersion delay in a single 96-kHz channel at the lowest frequency. The error bar in each burst shown in Figure~\ref{fig:diagnostics} is taken as the half-width at half-maximum of a Gaussian of appropriate amplitude. 


In addition, the bandpass-normalized dynamic spectrum shows scintillation, evident from the peaks in the $t<0$ and $t>0$ spectra of burst 19: we measure a scintillation bandwidth of 0.1-0.2~MHz at 550-750~MHz. Some of these features appear to be blended at the lower part of the band, whereas they appear to be adequately resolved $\gtrsim$620~MHz. At the frequencies at which we observe, the scattering tail appears to be scintillating with a characteristic bandwidth of $\sim$0.1~MHz but barely resolved: this might be an overestimate as the features appear to be blended. We hence have a case analogous to that of \cite{masui+15} where the scattering is thought to occur in the host galaxy, while the scatter-broadened burst still appears ``unresolved'' at a screen in the Milky way, as the scintillation appears constant throughout the scattering tail. We present a more detailed analysis of the scattering and scintillation in P-II, combining the measurements obtained from these Band-4 uGMRT and 100-m Effelsberg Radio Telescope L-band observations. 

It is necessary to distinguish the scattering time constant from the scintillation bandwidth. In our case, the scattering time constant is not inversely related to the measured scintillation bandwidth, leading to our hypothesis that the two measurements are distinct and the two phenomena are mutually independent. In fact, the scattering time constant is roughly 2 orders of magnitude larger than that expected from the scintillation bandwidth. P-II develops the hypothesis that the scattering arises from a screen that is located much closer to the FRB, possibly within the host galaxy, and that the scintillation is being imparted by a scattering screen in the Milky Way.

Importantly, the fact that both scattering and scintillation are seen in this FRB allows us to place constraints on the locations of the screens and measure a velocity of the Milky Way scattering screen. The inferred scattering screen transverse velocity is $30$-$40\, \mathrm{km\,s^{-1}}$ for a screen located at 2~kpc, and much smaller for a 0.4~kpc screen, when the velocity of the scintillation pattern has a larger contribution from Earth's velocity. For a screen located so close to Earth, its large contribution to the scintillation velocity is expected to manifest as a strong annual modulation of the scintillation timescale. If the scattering arises at the host galaxy, we expect the scattering timescale to be constant throughout the annual scintillation modulation cycle.

\subsection{Periodicity search on small timescales}

A {\tt PRESTO} search for periodicities on short timescales was done with acceleration and jerk on the full dedispersed data, as well as on multiple 10-minute and 20-minute segments. No significant candidates were detected. Additionally, a Fast Folding Algorithm (FFA) search was performed on the IA beam data using {\tt riptide}\footnote{\href{https://github.com/v-morello/riptide}{https://github.com/v-morello/riptide}} (\cite{Morello20}). However, that again revealed no significant candidates. Some bursts show multiple peaks, as in bursts 2, 13 and 22. It is not clear if these are distinct components of the same burst or are different bursts, due to the poor S/N of the IA beam. Their separations appear to be $\sim$ 90~ms or $\sim$180~ms. We verified through an autocorrelation analysis that the separations have no periodic relationship between the bursts. 

The detection of periodicity in the range of typical pulsar periods in FRBs would be unassailable evidence for a neutron star origin for the emission and repetition. While longer periodicites of $\sim$ several days, such as known for \riii\ \citep{chime2020a} relies on models like orbital motion \citep{16DaiOrbit, 20IokaOrbit} or precession \citep{20LevinPrecession}, smaller periodicities in the 1~ms to 1~s range are thought to arise from rotation. \cite{chime-subsecond-periodicity-2021} have discovered multiple bursts which appear clustered with sub-second separations, from FRB~20191221A ($\sim217$~ms), FRB~20210206A ($\sim3$~ms)  and FRB~20210213A ($\sim11$~ms), consistent with typical periods expected for magnetospheric emission. This provides strong motivation for follow-up observations of active repeaters like \rss\ with longer exposures, specifically  with the ability to track the source in the sky. 
Observations such as those described here have the potential to detect many more such clustered bursts. 

In this context it is interesting to note the possibility of ultralong period (ULP) magnetars as FRB candidates. \cite{Beniamini2020a} invoke a simple long rotation period explanation for the $\sim$16-day periodicity of \riii. In their model, ULP magnetars arise from three distinct possibilities, such as mass-loaded charged-particle winds, loss of angular momentum due to kicks from giant flares and long-lasting accretion discs \citep{Beniamini2020a}, citing the example of the Galactic magnetar candidate with a $\sim$7h period, 1E~161348–5055. However, they note that such progenitors would be extremely rare, accounting for the paucity of scales of periodicity similar to \riii.





\section{SUMMARY AND CONCLUSIONS}\label{sec:summary}
We report the observations at Band-4 of \rss\ with the upgraded GMRT and the detection of 48 bursts in 3 hours of exposure. These observations were carried out with the dual purpose of localizing the FRB as well as detecting associated persistent emission, based on the initial, non-interferometric localization reported by ASKAP \citep{ASKAP-secondATel}. This necessitated only a single pointing with the GMRT. We localize the FRB and detect persistent emission at 650~MHz.

We estimate the DM that maximizes the frequency-averaged substructure energy to be $410.78\pm0.54$\dm, similar to that obtained from {\tt DM\_phase}, $410.33\pm1.15$\dm. Both these DMs are consistent with the fiducial value of 411\dm\ used throughout this paper. A larger number of high S/N bursts from high sensitivity observations with the PA beam are essential to constrain any burst-to-burst DM variability within the observation, or for searching for systematic trends in DM variability over longer timescales \citep[e.g.][seen in \rAO]{Hilmarsson2021}. We find that the mean burst width is $17.5\pm7.2$~ms, but find that they are larger than the widths for other known repeaters. The isotropic equivalent burst spectral energies are $\sim10^{29}\mathrm{-}10^{31} $~erg\,Hz$^{-1}$.

The bursts range in fluence from 2.6\flu\ to 108\flu, following a power law distribution with an index of $\gamma=-1.2\pm0.2$ and an overall burst rate of 16 per hour for these observations, which is a lower limit. We estimate a completeness limit of 10\flu\ for these observations, after empirically modelling the burst width distribution. Our observations might have missed $\gtrsim$80\% of the bursts in the fluence range 1-10\flu. The burst waiting time distribution agrees broadly with an exponential distribution with a mean waiting time of $t_w \sim 2.9$~min, suggesting an underlying Poisson point process that is expected for bursts occurring within a short observation. There is no evidence for longer (minute to several minutes) periodicity or, more specifically, burst rate modulation with those characteristic timescales, inferred from the distribution of pairwise burst intervals in our observations. More sensitive observations with a tighter fluence completeness limit might revise these statistics considerably. 

With a DM=411\dm, we measure a range of bulk spectro-temporal drift rates between $-0.75$ and $-20~{\rm MHz~ms}^{-1}$. We measure an upper limit to the scattering time of $\leq$11.1~ms, as well as detect scintillation across frequency.
No small timescale, pulsar-like periodicity was detected in acceleration and jerk searches as well as with FFA.

The simultaneous localization of \rss\ adds strength to the proof-of-concept method adopted from our earlier work \citep{marthi+20}. This could serve as a potential model for all our future localization and follow up of unlocalized repeating FRBs.

More sensitive PA beam observations will likely yield an order of magnitude higher burst rate, if \rss\ continues in its present active state. This holds enormous promise for detailed studies of its polarization properties that would allow us to test between competing emission models, as well as study any systematic evolution of RM as reported in \cite{Hilmarsson2021}. Regular long-term monitoring would enable us to construct a putative trend for an annual scintillation timescale modulation, and hence to locate and study the Galactic scattering screen responsible for the scintillation. Variation or modulation of burst rate or activity over longer timescales would have to be adequately sampled to search for evidence for clustering, as is known for \rAO. Finally, although we have reported only radio observations, extremely active, localized repeating FRBs such as \rss\ are the best targets for extended multi-wavelength campaigns that can considerably advance our current understanding of the origins of FRBs.

\section{ACKNOWLEDGEMENTS}
We thank the anonymous reviewer for comments which have helped improve the manuscript. VRM thanks Jayaram Chengalur and Apurba Bera for insightful discussions. We thank the staff of the GMRT who have made these observations possible. GMRT is run by the National Centre for Radio Astrophysics of the Tata Institute of Fundamental Research. VRM acknowledges the support of the Department of Atomic Energy, Government of India, under project no. 12-R\&D-TFR-5.02-0700. We acknowledge use of the CHIME/FRB Public Database, provided at \url{https://www.chime-frb.ca/} by the CHIME/FRB Collaboration. We acknowledge the support of the Natural Sciences and Engineering Research Council of Canada (NSERC) (funding reference number RGPIN-2019-067, CRD 523638-201). We receive support from Ontario Research Fund - Research Excellence Program (ORF-RE), Canadian Institute for Advanced Research (CIFAR), Canadian Foundation for Innovation (CFI), Simons Foundation, Thoth Technology Inc. and Alexander von Humboldt Foundation. 
Part of this research was carried out at the Jet Propulsion Laboratory, California Institute of Technology, under a contract with the National Aeronautics and Space Administration. LGS is a Lise Meitner Indepdendent Max Planck reasearch group leader and acknowledges support from the Max Planck Society. 

\section{DATA AVAILABILITY}
The data underlying this article will be shared on reasonable request to the corresponding authors.

\bibliographystyle{mnras}
\bibliography{frb20201124a}

\begin{thebibliography}{}
\makeatletter
\relax
\def\mn@urlcharsother{\let\do\@makeother \do\$\do\&\do\#\do\^\do\_\do\%\do\~}
\def\mn@doi{\begingroup\mn@urlcharsother \@ifnextchar [ {\mn@doi@}
  {\mn@doi@[]}}
\def\mn@doi@[#1]#2{\def\@tempa{#1}\ifx\@tempa\@empty \href
  {http://dx.doi.org/#2} {doi:#2}\else \href {http://dx.doi.org/#2} {#1}\fi
  \endgroup}
\def\mn@eprint#1#2{\mn@eprint@#1:#2::\@nil}
\def\mn@eprint@arXiv#1{\href {http://arxiv.org/abs/#1} {{\tt arXiv:#1}}}
\def\mn@eprint@dblp#1{\href {http://dblp.uni-trier.de/rec/bibtex/#1.xml}
  {dblp:#1}}
\def\mn@eprint@#1:#2:#3:#4\@nil{\def\@tempa {#1}\def\@tempb {#2}\def\@tempc
  {#3}\ifx \@tempc \@empty \let \@tempc \@tempb \let \@tempb \@tempa \fi \ifx
  \@tempb \@empty \def\@tempb {arXiv}\fi \@ifundefined
  {mn@eprint@\@tempb}{\@tempb:\@tempc}{\expandafter \expandafter \csname
  mn@eprint@\@tempb\endcsname \expandafter{\@tempc}}}

\bibitem[\protect\citeauthoryear{{Beniamini}, {Wadiasingh}  \&
  {Metzger}}{{Beniamini} et~al.}{2020}]{Beniamini2020a}
{Beniamini} P.,  {Wadiasingh} Z.,   {Metzger} B.~D.,  2020, \mn@doi [\mnras]
  {10.1093/mnras/staa1783}, \href
  {https://ui.adsabs.harvard.edu/abs/2020MNRAS.496.3390B} {496, 3390}

\bibitem[\protect\citeauthoryear{{Bera} \& {Chengalur}}{{Bera} \&
  {Chengalur}}{2019}]{Bera2019}
{Bera} A.,  {Chengalur} J.~N.,  2019, \mn@doi [\mnras] {10.1093/mnrasl/slz140},
  \href {https://ui.adsabs.harvard.edu/abs/2019MNRAS.490L..12B} {490, L12}

\bibitem[\protect\citeauthoryear{{Bhardwaj} et~al.,}{{Bhardwaj}
  et~al.}{2021}]{Bhardwaj2021}
{Bhardwaj} M.,  et~al., 2021, \mn@doi [\apjl] {10.3847/2041-8213/abeaa6}, \href
  {https://ui.adsabs.harvard.edu/abs/2021ApJ...910L..18B} {910, L18}

\bibitem[\protect\citeauthoryear{{Bochenek}, {Ravi}, {Belov}, {Hallinan},
  {Kocz}, {Kulkarni}  \& {McKenna}}{{Bochenek} et~al.}{2020}]{bochenek+20}
{Bochenek} C.~D.,  {Ravi} V.,  {Belov} K.~V.,  {Hallinan} G.,  {Kocz} J.,
  {Kulkarni} S.~R.,   {McKenna} D.~L.,  2020, arXiv e-prints, \href
  {https://ui.adsabs.harvard.edu/abs/2020arXiv200510828B} {p. arXiv:2005.10828}

\bibitem[\protect\citeauthoryear{{CHIME/FRB Collaboration}}{{CHIME/FRB
  Collaboration}}{2021}]{CHIMER67ATel}
{CHIME/FRB Collaboration} 2021, The Astronomer's Telegram, \href
  {https://ui.adsabs.harvard.edu/abs/2021ATel14497....1C} {14497, 1}

\bibitem[\protect\citeauthoryear{{CHIME/FRB Collaboration} et~al.,}{{CHIME/FRB
  Collaboration} et~al.}{2020a}]{chime2020a}
{CHIME/FRB Collaboration} et~al., 2020a, \mn@doi [\nat]
  {10.1038/s41586-020-2398-2}, \href
  {https://ui.adsabs.harvard.edu/abs/2020Natur.582..351C} {582, 351}

\bibitem[\protect\citeauthoryear{{CHIME/FRB Collaboration} et~al.,}{{CHIME/FRB
  Collaboration} et~al.}{2020b}]{chime2020b}
{CHIME/FRB Collaboration} et~al., 2020b, \mn@doi [\nat]
  {10.1038/s41586-020-2863-y}, \href
  {https://ui.adsabs.harvard.edu/abs/2020Natur.587...54C} {587, 54}

\bibitem[\protect\citeauthoryear{{Caleb} et~al.,}{{Caleb}
  et~al.}{2020}]{Caleb2020}
{Caleb} M.,  et~al., 2020, \mn@doi [\mnras] {10.1093/mnras/staa1791}, \href
  {https://ui.adsabs.harvard.edu/abs/2020MNRAS.496.4565C} {496, 4565}

\bibitem[\protect\citeauthoryear{{Chamma}, {Rajabi}, {Wyenberg}, {Mathews}  \&
  {Houde}}{{Chamma} et~al.}{2020}]{Chamma2020arXiv}
{Chamma} M.~A.,  {Rajabi} F.,  {Wyenberg} C.~M.,  {Mathews} A.,   {Houde} M.,
  2020, arXiv e-prints, \href
  {https://ui.adsabs.harvard.edu/abs/2020arXiv201014041C} {p. arXiv:2010.14041}

\bibitem[\protect\citeauthoryear{{Chawla} et~al.,}{{Chawla}
  et~al.}{2020}]{Chawla2020}
{Chawla} P.,  et~al., 2020, \mn@doi [\apjl] {10.3847/2041-8213/ab96bf}, \href
  {https://ui.adsabs.harvard.edu/abs/2020ApJ...896L..41C} {896, L41}

\bibitem[\protect\citeauthoryear{{Cruces} et~al.,}{{Cruces}
  et~al.}{2021}]{Cruces+2021}
{Cruces} M.,  et~al., 2021, \mn@doi [\mnras] {10.1093/mnras/staa3223}, \href
  {https://ui.adsabs.harvard.edu/abs/2021MNRAS.500..448C} {500, 448}

\bibitem[\protect\citeauthoryear{{Dai}, {Wang}, {Wu}  \& {Huang}}{{Dai}
  et~al.}{2016}]{16DaiOrbit}
{Dai} Z.~G.,  {Wang} J.~S.,  {Wu} X.~F.,   {Huang} Y.~F.,  2016, \mn@doi [\apj]
  {10.3847/0004-637X/829/1/27}, \href
  {https://ui.adsabs.harvard.edu/abs/2016ApJ...829...27D} {829, 27}

\bibitem[\protect\citeauthoryear{{Day}, {Bhandari}, {Deller}, {Shannon}  \&
  {Moss}}{{Day} et~al.}{2021a}]{ASKAP-localization-ATel}
{Day} C.~K.,  {Bhandari} S.,  {Deller} A.~T.,  {Shannon} R.~M.,   {Moss} V.~A.,
   2021a, The Astronomer's Telegram, \href
  {https://ui.adsabs.harvard.edu/abs/2021ATel14515....1D} {14515, 1}

\bibitem[\protect\citeauthoryear{{Day}, {Bhandari}, {Deller}, {Shannon}  \&
  {ASKAP-CRAFT Survey Science Project}}{{Day}
  et~al.}{2021b}]{ASKAP-lowband-loc-ATel}
{Day} C.~K.,  {Bhandari} S.,  {Deller} A.~T.,  {Shannon} R.~M.,   {ASKAP-CRAFT
  Survey Science Project} 2021b, The Astronomer's Telegram, \href
  {https://ui.adsabs.harvard.edu/abs/2021ATel14592....1D} {14592, 1}

\bibitem[\protect\citeauthoryear{{Fong} et~al.,}{{Fong}
  et~al.}{2021}]{Fong2021arXiv}
{Fong} W.-f.,  et~al., 2021, arXiv e-prints, \href
  {https://ui.adsabs.harvard.edu/abs/2021arXiv210611993F} {p. arXiv:2106.11993}

\bibitem[\protect\citeauthoryear{{Gourdji}, {Michilli}, {Spitler}, {Hessels},
  {Seymour}, {Cordes}  \& {Chatterjee}}{{Gourdji} et~al.}{2019}]{Gourdji2019}
{Gourdji} K.,  {Michilli} D.,  {Spitler} L.~G.,  {Hessels} J.~W.~T.,  {Seymour}
  A.,  {Cordes} J.~M.,   {Chatterjee} S.,  2019, \mn@doi [\apjl]
  {10.3847/2041-8213/ab1f8a}, \href
  {https://ui.adsabs.harvard.edu/abs/2019ApJ...877L..19G} {877, L19}

\bibitem[\protect\citeauthoryear{{Gupta} et~al.,}{{Gupta}
  et~al.}{2017}]{uGMRTpaper}
{Gupta} Y.,  et~al., 2017, Current Science, \href
  {https://ui.adsabs.harvard.edu/abs/2017CSci..113..707G} {113, 707}

\bibitem[\protect\citeauthoryear{{Hessels} et~al.,}{{Hessels}
  et~al.}{2019}]{hessels+19}
{Hessels} J.~W.~T.,  et~al., 2019, \mn@doi [\apjl] {10.3847/2041-8213/ab13ae},
  \href {https://ui.adsabs.harvard.edu/abs/2019ApJ...876L..23H} {876, L23}

\bibitem[\protect\citeauthoryear{{Hilmarsson}, {Spitler}, {Main}  \&
  {Li}}{{Hilmarsson} et~al.}{2021a}]{Hilmarsson2021arXiv}
{Hilmarsson} G.~H.,  {Spitler} L.~G.,  {Main} R.~A.,   {Li} D.~Z.,  2021a,
  arXiv e-prints, \href {https://ui.adsabs.harvard.edu/abs/2021arXiv210712892H}
  {p. arXiv:2107.12892}

\bibitem[\protect\citeauthoryear{{Hilmarsson} et~al.,}{{Hilmarsson}
  et~al.}{2021b}]{Hilmarsson2021}
{Hilmarsson} G.~H.,  et~al., 2021b, \mn@doi [\apjl] {10.3847/2041-8213/abdec0},
  \href {https://ui.adsabs.harvard.edu/abs/2021ApJ...908L..10H} {908, L10}

\bibitem[\protect\citeauthoryear{{Ioka} \& {Zhang}}{{Ioka} \&
  {Zhang}}{2020}]{20IokaOrbit}
{Ioka} K.,  {Zhang} B.,  2020, \mn@doi [\apjl] {10.3847/2041-8213/ab83fb},
  \href {https://ui.adsabs.harvard.edu/abs/2020ApJ...893L..26I} {893, L26}

\bibitem[\protect\citeauthoryear{{James}, {Ekers}, {Macquart}, {Bannister}  \&
  {Shannon}}{{James} et~al.}{2019}]{james+2019}
{James} C.~W.,  {Ekers} R.~D.,  {Macquart} J.~P.,  {Bannister} K.~W.,
  {Shannon} R.~M.,  2019, \mn@doi [\mnras] {10.1093/mnras/sty3031}, \href
  {https://ui.adsabs.harvard.edu/abs/2019MNRAS.483.1342J} {483, 1342}

\bibitem[\protect\citeauthoryear{{Karuppusamy}, {Stappers}  \& {van
  Straten}}{{Karuppusamy} et~al.}{2010}]{ramesh10}
{Karuppusamy} R.,  {Stappers} B.~W.,   {van Straten} W.,  2010, \mn@doi [\aap]
  {10.1051/0004-6361/200913729}, \href
  {https://ui.adsabs.harvard.edu/abs/2010A&A...515A..36K} {515, A36}

\bibitem[\protect\citeauthoryear{{Kilpatrick}, {Fong}, {Prochaska}, {Tejos},
  {Bhandari}  \& {Day}}{{Kilpatrick} et~al.}{2021}]{R67-redshift-Atel}
{Kilpatrick} C.~D.,  {Fong} W.,  {Prochaska} J.~X.,  {Tejos} N.,  {Bhandari}
  S.,   {Day} C.~K.,  2021, The Astronomer's Telegram, \href
  {https://ui.adsabs.harvard.edu/abs/2021ATel14516....1K} {14516, 1}

\bibitem[\protect\citeauthoryear{{Kirsten} et~al.,}{{Kirsten}
  et~al.}{2021}]{Kirsten2021arXiv}
{Kirsten} F.,  et~al., 2021, arXiv e-prints, \href
  {https://ui.adsabs.harvard.edu/abs/2021arXiv210511445K} {p. arXiv:2105.11445}

\bibitem[\protect\citeauthoryear{{Kumar}, {Shannon}, {Moss}, {Qiu}  \&
  {Bhandari}}{{Kumar} et~al.}{2021a}]{ASKAP-firstATel}
{Kumar} P.,  {Shannon} R.~M.,  {Moss} V.,  {Qiu} H.,   {Bhandari} S.,  2021a,
  The Astronomer's Telegram, \href
  {https://ui.adsabs.harvard.edu/abs/2021ATel14502....1K} {14502, 1}

\bibitem[\protect\citeauthoryear{{Kumar}, {Shannon}, {Keane}, {Moss}  \&
  {Askap-Craft Survey Science Project}}{{Kumar}
  et~al.}{2021b}]{ASKAP-secondATel}
{Kumar} P.,  {Shannon} R.~M.,  {Keane} E.,  {Moss} V.~A.,   {Askap-Craft Survey
  Science Project} 2021b, The Astronomer's Telegram, \href
  {https://ui.adsabs.harvard.edu/abs/2021ATel14508....1K} {14508, 1}

\bibitem[\protect\citeauthoryear{{Law}, {Tendulkar}, {Clarke}, {Aggarwal}  \&
  {Bethapudy}}{{Law} et~al.}{2021}]{VLA-localization-ATel}
{Law} C.,  {Tendulkar} S.,  {Clarke} T.,  {Aggarwal} K.,   {Bethapudy} S.,
  2021, The Astronomer's Telegram, \href
  {https://ui.adsabs.harvard.edu/abs/2021ATel14526....1L} {14526, 1}

\bibitem[\protect\citeauthoryear{{Levin}, {Beloborodov}  \&
  {Bransgrove}}{{Levin} et~al.}{2020}]{20LevinPrecession}
{Levin} Y.,  {Beloborodov} A.~M.,   {Bransgrove} A.,  2020, \mn@doi [\apjl]
  {10.3847/2041-8213/ab8c4c}, \href
  {https://ui.adsabs.harvard.edu/abs/2020ApJ...895L..30L} {895, L30}

\bibitem[\protect\citeauthoryear{{Li} et~al.,}{{Li} et~al.}{2021}]{Li+2021}
{Li} D.,  et~al., 2021, arXiv e-prints, \href
  {https://ui.adsabs.harvard.edu/abs/2021arXiv210708205L} {p. arXiv:2107.08205}

\bibitem[\protect\citeauthoryear{{Main}, {Hilmarsson}, {Marthi}, {Spitler},
  {Wharton}, {Bethapudi}, {Li}  \& {Lin}}{{Main} et~al.}{2021}]{main+2021}
{Main} R.~A.,  {Hilmarsson} G.~H.,  {Marthi} V.~R.,  {Spitler} L.~G.,
  {Wharton} R.~S.,  {Bethapudi} S.,  {Li} D.~Z.,   {Lin} H.~H.,  2021, arXiv
  e-prints, \href {https://ui.adsabs.harvard.edu/abs/2021arXiv210800052M} {p.
  arXiv:2108.00052}

\bibitem[\protect\citeauthoryear{{Majid} et~al.,}{{Majid}
  et~al.}{2021}]{Majid+2021}
{Majid} W.~A.,  et~al., 2021, arXiv e-prints, \href
  {https://ui.adsabs.harvard.edu/abs/2021arXiv210510987M} {p. arXiv:2105.10987}

\bibitem[\protect\citeauthoryear{{Marcote} et~al.,}{{Marcote}
  et~al.}{2021}]{EVN-localization-ATel}
{Marcote} B.,  et~al., 2021, The Astronomer's Telegram, \href
  {https://ui.adsabs.harvard.edu/abs/2021ATel14603....1M} {14603, 1}

\bibitem[\protect\citeauthoryear{{Margalit} \& {Metzger}}{{Margalit} \&
  {Metzger}}{2018}]{Margalit2018}
{Margalit} B.,  {Metzger} B.~D.,  2018, \mn@doi [\apjl]
  {10.3847/2041-8213/aaedad}, \href
  {https://ui.adsabs.harvard.edu/abs/2018ApJ...868L...4M} {868, L4}

\bibitem[\protect\citeauthoryear{{Marthi}, {Gautam}, {Li}, {Lin}, {Main},
  {Naidu}, {Pen}  \& {Wharton}}{{Marthi} et~al.}{2020}]{marthi+20}
{Marthi} V.~R.,  {Gautam} T.,  {Li} D.~Z.,  {Lin} H.~H.,  {Main} R.~A.,
  {Naidu} A.,  {Pen} U.~L.,   {Wharton} R.~S.,  2020, \mn@doi [\mnras]
  {10.1093/mnrasl/slaa148}, \href
  {https://ui.adsabs.harvard.edu/abs/2020MNRAS.499L..16M} {499, L16}

\bibitem[\protect\citeauthoryear{{Masui} et~al.,}{{Masui}
  et~al.}{2015}]{masui+15}
{Masui} K.,  et~al., 2015, \mn@doi [\nat] {10.1038/nature15769}, \href
  {https://ui.adsabs.harvard.edu/abs/2015Natur.528..523M} {528, 523}

\bibitem[\protect\citeauthoryear{{Morello}, {Barr}, {Stappers}, {Keane}  \&
  {Lyne}}{{Morello} et~al.}{2020}]{Morello20}
{Morello} V.,  {Barr} E.~D.,  {Stappers} B.~W.,  {Keane} E.~F.,   {Lyne} A.~G.,
   2020, \mn@doi [\mnras] {10.1093/mnras/staa2291}, \href
  {https://ui.adsabs.harvard.edu/abs/2020MNRAS.497.4654M} {497, 4654}

\bibitem[\protect\citeauthoryear{{Nimmo} et~al.,}{{Nimmo}
  et~al.}{2021}]{Nimmo+2021}
{Nimmo} K.,  et~al., 2021, arXiv e-prints, \href
  {https://ui.adsabs.harvard.edu/abs/2021arXiv210511446N} {p. arXiv:2105.11446}

\bibitem[\protect\citeauthoryear{{Oppermann}, {Yu}  \& {Pen}}{{Oppermann}
  et~al.}{2018}]{Oppermann2018}
{Oppermann} N.,  {Yu} H.-R.,   {Pen} U.-L.,  2018, \mn@doi [\mnras]
  {10.1093/mnras/sty004}, \href
  {https://ui.adsabs.harvard.edu/abs/2018MNRAS.475.5109O} {475, 5109}

\bibitem[\protect\citeauthoryear{{Perley} \& {Butler}}{{Perley} \&
  {Butler}}{2017}]{Perley2017}
{Perley} R.~A.,  {Butler} B.~J.,  2017, \mn@doi [\apjs]
  {10.3847/1538-4365/aa6df9}, \href
  {https://ui.adsabs.harvard.edu/abs/2017ApJS..230....7P} {230, 7}

\bibitem[\protect\citeauthoryear{{Piro} \& {Gaensler}}{{Piro} \&
  {Gaensler}}{2018}]{Piro2018}
{Piro} A.~L.,  {Gaensler} B.~M.,  2018, \mn@doi [\apj]
  {10.3847/1538-4357/aac9bc}, \href
  {https://ui.adsabs.harvard.edu/abs/2018ApJ...861..150P} {861, 150}

\bibitem[\protect\citeauthoryear{{Platts}, {Weltman}, {Walters}, {Tendulkar},
  {Gordin}  \& {Kandhai}}{{Platts} et~al.}{2019}]{Platts2019}
{Platts} E.,  {Weltman} A.,  {Walters} A.,  {Tendulkar} S.~P.,  {Gordin}
  J.~E.~B.,   {Kandhai} S.,  2019, \mn@doi [\physrep]
  {10.1016/j.physrep.2019.06.003}, \href
  {https://ui.adsabs.harvard.edu/abs/2019PhR...821....1P} {821, 1}

\bibitem[\protect\citeauthoryear{{Platts} et~al.,}{{Platts}
  et~al.}{2021}]{Platts2021}
{Platts} E.,  et~al., 2021, \mn@doi [\mnras] {10.1093/mnras/stab1544}, \href
  {https://ui.adsabs.harvard.edu/abs/2021MNRAS.505.3041P} {505, 3041}

\bibitem[\protect\citeauthoryear{{Pleunis} et~al.,}{{Pleunis}
  et~al.}{2021a}]{2021arXiv210604356P}
{Pleunis} Z.,  et~al., 2021a, arXiv e-prints, \href
  {https://ui.adsabs.harvard.edu/abs/2021arXiv210604356P} {p. arXiv:2106.04356}

\bibitem[\protect\citeauthoryear{{Pleunis} et~al.,}{{Pleunis}
  et~al.}{2021b}]{Pleunis2021}
{Pleunis} Z.,  et~al., 2021b, \mn@doi [\apjl] {10.3847/2041-8213/abec72}, \href
  {https://ui.adsabs.harvard.edu/abs/2021ApJ...911L...3P} {911, L3}

\bibitem[\protect\citeauthoryear{{Rajwade} et~al.,}{{Rajwade}
  et~al.}{2020}]{20Rajwade}
{Rajwade} K.~M.,  et~al., 2020, \mn@doi [\mnras] {10.1093/mnras/staa1237},
  \href {https://ui.adsabs.harvard.edu/abs/2020MNRAS.495.3551R} {495, 3551}

\bibitem[\protect\citeauthoryear{{Ransom}, {Eikenberry}  \&
  {Middleditch}}{{Ransom} et~al.}{2002}]{Ransom2002}
{Ransom} S.~M.,  {Eikenberry} S.~S.,   {Middleditch} J.,  2002, \mn@doi [\aj]
  {10.1086/342285}, \href
  {https://ui.adsabs.harvard.edu/abs/2002AJ....124.1788R} {124, 1788}

\bibitem[\protect\citeauthoryear{{Ravi} et~al.,}{{Ravi}
  et~al.}{2021}]{Ravi2021arXiv}
{Ravi} V.,  et~al., 2021, arXiv e-prints, \href
  {https://ui.adsabs.harvard.edu/abs/2021arXiv210609710R} {p. arXiv:2106.09710}

\bibitem[\protect\citeauthoryear{{Scholz} et~al.,}{{Scholz}
  et~al.}{2016}]{scholz2016}
{Scholz} P.,  et~al., 2016, \mn@doi [ApJ] {10.3847/1538-4357/833/2/177}, \href
  {https://ui.adsabs.harvard.edu/abs/2016ApJ...833..177S} {833, 177}

\bibitem[\protect\citeauthoryear{{Seymour}, {Michilli}  \& {Pleunis}}{{Seymour}
  et~al.}{2019}]{DM_phase_pub}
{Seymour} A.,  {Michilli} D.,   {Pleunis} Z.,  2019, {DM\_phase: Algorithm for
  correcting dispersion of radio signals} (\mn@eprint {ascl} {1910.004})

\bibitem[\protect\citeauthoryear{{Spitler} et~al.,}{{Spitler}
  et~al.}{2016}]{Spitler2016}
{Spitler} L.~G.,  et~al., 2016, \mn@doi [\nat] {10.1038/nature17168}, \href
  {https://ui.adsabs.harvard.edu/abs/2016Natur.531..202S} {531, 202}

\bibitem[\protect\citeauthoryear{{The CHIME/FRB Collaboration} et~al.,}{{The
  CHIME/FRB Collaboration} et~al.}{2020}]{20R3Period}
{The CHIME/FRB Collaboration} et~al., 2020, arXiv e-prints, \href
  {https://ui.adsabs.harvard.edu/abs/2020arXiv200110275T} {p. arXiv:2001.10275}

\bibitem[\protect\citeauthoryear{{The CHIME/FRB Collaboration} et~al.,}{{The
  CHIME/FRB Collaboration} et~al.}{2021}]{chime-subsecond-periodicity-2021}
{The CHIME/FRB Collaboration} et~al., 2021, arXiv e-prints, \href
  {https://ui.adsabs.harvard.edu/abs/2021arXiv210708463T} {p. arXiv:2107.08463}

\bibitem[\protect\citeauthoryear{{Thompson}}{{Thompson}}{2017}]{Thompson2017}
{Thompson} C.,  2017, \mn@doi [\apj] {10.3847/1538-4357/aa7845}, \href
  {https://ui.adsabs.harvard.edu/abs/2017ApJ...844..162T} {844, 162}

\bibitem[\protect\citeauthoryear{{Thompson}}{{Thompson}}{2019}]{Thompson2019}
{Thompson} C.,  2019, \mn@doi [\apj] {10.3847/1538-4357/aafda3}, \href
  {https://ui.adsabs.harvard.edu/abs/2019ApJ...874...48T} {874, 48}

\bibitem[\protect\citeauthoryear{{Wharton} et~al.,}{{Wharton}
  et~al.}{2021a}]{GMRT-PRS-ATel}
{Wharton} R.,  et~al., 2021a, The Astronomer's Telegram, \href
  {https://ui.adsabs.harvard.edu/abs/2021ATel14529....1W} {14529, 1}

\bibitem[\protect\citeauthoryear{{Wharton} et~al.,}{{Wharton}
  et~al.}{2021b}]{GMRT-localization-ATel}
{Wharton} R.,  et~al., 2021b, The Astronomer's Telegram, \href
  {https://ui.adsabs.harvard.edu/abs/2021ATel14538....1W} {14538, 1}

\bibitem[\protect\citeauthoryear{{Zhang}, {Gajjar}, {Foster}, {Siemion},
  {Cordes}, {Law}  \& {Wang}}{{Zhang} et~al.}{2018}]{Zhang2018}
{Zhang} Y.~G.,  {Gajjar} V.,  {Foster} G.,  {Siemion} A.,  {Cordes} J.,  {Law}
  C.,   {Wang} Y.,  2018, \mn@doi [\apj] {10.3847/1538-4357/aadf31}, \href
  {https://ui.adsabs.harvard.edu/abs/2018ApJ...866..149Z} {866, 149}

\makeatother
\end{thebibliography}

\end{document}